\documentclass{jfm}

\usepackage{natbib}
\usepackage{url}

\usepackage{amsmath,mathtools}
\usepackage{amssymb,bm}
\usepackage{array}
\usepackage{tikz}
\usepackage{subcaption}
\newcommand{\avg}[1]{\left< #1 \right>}     
\newcommand{\vect}[1]{\bm{#1}}          
\newcommand{\ex}{\vect{e}_x}            
\newcommand{\dd}{\,\mathrm{d}}          
\newcommand{\defeq}{\coloneqq}          

\newcommand{\lb}[1]{\left(}
\newcommand{\rb}[1]{\right)}
\newcommand{\on}[1]{\,\text{on } #1}
\newcommand{\eqrefrangecompact}[2]{%
  (\ref{#1}--\ref{#2})%
}
\newcommand{\gridpaneltags}{%
  \begin{scope}[x={(gridimage.south east)},y={(gridimage.north west)}]
    \tikzset{paneltag/.style={inner sep=0pt,font=\small}}
    \node[paneltag,anchor=south west] at (0.111,0.985) {(a)};
    \node[paneltag,anchor=south west] at (0.111,0.495) {(b)};
    \node[paneltag,anchor=south west] at (0.405,0.495) {(c)};
    \node[paneltag,anchor=south west] at (0.700,0.495) {(d)};
    \node[paneltag,anchor=south west] at (0.111,0.248) {(e)};
    \node[paneltag,anchor=south west] at (0.405,0.248) {(f)};
    \node[paneltag,anchor=south west] at (0.700,0.248) {(g)};
  \end{scope}%
}

\shorttitle{Wave-driven propulsion of a flexible raft}
\shortauthor{E. A. Agüero and D. M. Harris}
\title{Wave-driven propulsion of a flexible raft}

\author{Elvis A. Agüero and Daniel M. Harris \corresp{\email{daniel\_harris3@brown.edu}}}

\affiliation{School of Engineering, Brown University, Providence, RI 02912, USA}

\begin{document}
\maketitle

\begin{abstract}
Inertial propulsion in fluids generally arises from an unbalanced flux of momentum.  For the case of wave-driven propulsion, momentum is transported away from an oscillating raft in the form of self-excited surface waves.  While this mechanism has previously been analyzed for rigid rafts, the role of flexibility has yet to be investigated.  In this work, we
develop a fluid-structure interaction model for a periodically driven two-dimensional flexible raft resting at the free surface of a fluid.  The raft is modeled as an Euler--Bernoulli beam and is coupled to a weakly dissipative quasi-potential model of the fluid beneath. Varying the flexural stiffness and forcing position reveals new features associated with the introduction of flexibility, including the possibilities of thrust enhancement and reversal.  By projecting the raft response onto its free rigid-body and elastic modes, the fluid loading can be represented as a modal impedance, informing a computationally efficient reduced-order model. Analysis of the modal response and its symmetries facilitates physical interpretation of our key findings.  For a uniform raft, excitation of any single mode in isolation is incapable of producing a net thrust, and thus efficient wave propulsion requires a blend of interfering modes with appropriately coordinated amplitudes and phases.

\end{abstract}



\section{Introduction}

Wave-driven propulsion is a mode of interfacial locomotion in which a periodically driven floating body moves by generating an asymmetric field of surface waves \citep{LonguetHiggins1977,RohGharib2019,RheeEtAl2022,BenhamEtAl2022,HoPucciOzaHarris2023}. It represents one of the many natural locomotion mechanisms at fluid interfaces \citep{BushHu2006}; however, it remains relatively unexplored compared to other strategies \citep{HarrisBarotta2025}. Although the undulatory kinematics of a wave-propelling flexible sheet has recently been considered in experiment \citep{RenUcakYanSitti2024}, all other prior experimental and modeling studies have focused on rigid-body raft dynamics.  As such, the influence of body flexibility on wave-driven propulsion has yet to be systematically investigated.

The role of body deformation for propulsion in bulk fluids (i.e. away from the free surface) has been a central theme throughout classical and modern studies of swimming. For instance, Taylor's analysis of an undulating sheet demonstrated how traveling transverse waves can directly drive locomotion in a viscous fluid \citep{Taylor1951}, and potential-flow analyses of flexible waving plates have established how thrust and efficiency depend on deformation wavelength \citep{Wu1961}. Later, fish-like robotic swimmers as well as computational studies have explored the role of body flexibility in achieving propulsion \citep{TriantafyllouEtAl2000} with both theory and experiment demonstrating that thrust can be greatly enhanced at particular bending rigidities where resonance between the driving frequency and structural modes amplifies trailing-edge motion \citep{Alben2008,ParazEtAl2016,HooverEtAl2018,Smits2019,QuinnLauder2022}. Furthermore, robotic caudal-fin experiments confirmed thrust itself peaking at intermediate stiffnesses \citep{EspositoEtAl2012}, and simulations found that passive fin deformation can improve propulsive efficiency by reshaping how pressure loads are distributed over the surface \citep{KumarEtAl2025}.  Nevertheless, the relationship between the excitation of elastic structural modes and surface wave generation has yet to be explored in the context of interfacial propulsion, and represents the central focus of the present work. 

Related wave--structure interaction (WSI) problems wherein a floating elastic sheet or plate is coupled to a surrounding wave field through flexural body dynamics have been studied previously in the context of applications such as wave propagation under ice sheets \citep{SquireEtAl1995,GaoWangMilewski2019,PierceLiuYue2024,ledoudicMeasuringLocalMechanical2025} and wave-induced drag and wave-energy harvesting \citep{LonguetHiggins1977,falnes2015fundamental,DodeEtAl2022}. These WSI applications typically reside at meter scales or larger, where any role of capillarity on either the fluid or structural dynamics is negligible, although recent efforts have investigated elastic floaters at centimetric scales \citep{dhote2025flexible,herreman2026preferential}.  At even smaller scales, capillary stresses can become comparable to elastic stresses in thin plates and sheets, and their delicate interplay is often referred to as elastocapillarity \citep{ReisEtAl2010,DupratAristoffStone2011,TaroniVella2012,RivettiAntkowiak2013,BicoReyssatRoman2018}.  Nevertheless, most prior studies have focused on static or quasi-static regimes, neglecting inertial effects of either the structure or the fluid.  The application of interest herein requires fully inertial dynamics and capillary effects on both the fluid and solid, and thus integrates concepts from the historically disparate fields of WSI and elastocapillarity.

For the case of wave-driven propulsion, a natural conceptual framework is that of mean momentum fluxes and radiation stresses \citep{LonguetHigginsStewart1964,BenhamDevauchelleThomson2024,HarrisBarotta2025,ODonovanEtAl2025}. The concept of radiation stress, the excess wave-averaged flux of horizontal momentum, was developed to explain wave set-up and wave--current interactions, and was previously extended to explicitly incorporate surface tension effects \citep{LonguetHigginsStewart1964}. Net wave thrust may be interpreted as the imbalance of wave momentum flux radiated by the fore and aft by an oscillating body, yielding a drift that is consistent with global momentum conservation, as first demonstrated experimentally for a small wave-propelled craft by \citet{LonguetHiggins1977}. This momentum-flux viewpoint has been used in recent theoretical treatments of wave-driven locomotion of floating bodies \citep{BenhamDevauchelleThomson2024,BenhamEtAl2022} and in estimating thrust forces in experiments \citep{RohGharib2019,RheeEtAl2022,HoPucciOzaHarris2023}.

A simple modern realization of wave-driven propulsion is the \emph{SurferBot}, a small-scale interfacial robot that achieves wave-driven propulsion via vibrations generated by an onboard vibration motor \citep{RheeEtAl2022}. Inspired by this device, \citet{BenhamDevauchelleThomson2024,benham2025waveCorr} subsequently developed a modeling framework that coupled the equations of motion of a rigid floating body to a quasi-potential fluid model \citep{DiasDyachenkoZakharov2008}, with results for drift speed and thrust consistent with momentum conservation and the theory of \citet{LonguetHigginsStewart1964}, as well as the experimentally measured SurferBot dynamics.  Using their model, \citet{BenhamDevauchelleThomson2024} explored how the frequency and position of vibration motor influence the locomotion efficiency, motivating future experiments and more rigorous optimization studies \citep{ODonovanEtAl2025,odonovan2026}.  The model developed by \citet{BenhamDevauchelleThomson2024} serves as the foundation for the present work.

Here we develop a model for the wave-driven propulsion of a two-dimensional flexible raft, modeled as an Euler--Bernoulli beam oscillating vertically at a fluid interface, directly coupled to a weakly dissipative quasi-potential fluid model through pressure loading and elastic response (Section~\ref{sec:model_formulation}). Requiring only a few numerical solves of the full model to construct, we also develop a predictive reduced-order model by projecting the raft shape onto its corresponding free rigid-body and elastic modes, and representing the fluid loading as a complex modal impedance (Section~\ref{sec:modal-reduction}).  Results are presented in Section~\ref{sec:results}.  We first validate the rigid-body limit of our model against the SurferBot measurements of \citet{RheeEtAl2022}.  We then demonstrate how the mean thrust and propulsion efficiency depend sensitively on flexural stiffness and actuator position, which combine to select a specific set of structural modes dominating the response.  Our results reveal new features associated with raft flexibility, such as enhanced thrust due to the resonant excitation of structural modes and the possible reversal of propulsion direction. Section~\ref{sec:discussion} summarizes the conclusions, contextualizes the results within related works, and outlines future directions and open questions.  The full simulation code is documented and released in a publicly accessible open-access repository as part of this work.

\begin{table}
\centering
\caption{Notation used in the model formulation.}
\label{tab:notation}
\setlength{\tabcolsep}{1pt}
\begin{tabular*}{\textwidth}{@{\extracolsep{\fill}}p{0.3\textwidth} c p{0.28\textwidth} c@{}}
\hline
Quantity (units) & Symbol & Quantity (units) & Symbol \\ \hline
Fluid density (kg m$^{-3}$) & $\rho$ & Free-surface elevation (m) & $\eta(x,t)$ \\
Kinematic viscosity (m$^2$ s$^{-1}$) & $\nu$ & Velocity potential (m$^2$ s$^{-1}$) & $\phi(x,z,t)$ \\
Surface tension (N m$^{-1}$) & $\sigma$ & Fluid pressure (N m$^{-2}$) & $p(x,z,t)$ \\
Gravity (m s$^{-2}$) & $g$ & Bending moment (N m) & $M(x,t)$ \\
Fluid depth (m) & $H$ & Density ratio & $\Gamma=\rho L^2/\rho_R$ \\
Raft length (m) & $L$ & Froude number & $Fr=\sqrt{L\omega^2/g}$ \\
Raft width (m) & $d$ & Reynolds number & $Re=L^2\omega/\nu$ \\
Flexural rigidity (N m$^2$) & $EI(x)$ & Flexural parameter & $\kappa=EI/(\rho_RL^4\omega^2)$ \\
Raft density (kg m$^{-1}$) & $\rho_R(x)$ & Weber number & $We=\rho_RL\omega^2/\sigma$ \\
Vertical load (N m$^{-1}$) & $f(x,t)$ & Width ratio & $\Lambda=d/L$ \\
Forcing frequency (s$^{-1}$) & $\omega$ & Real dimensional variable & $\phi$ \\
Mean horizontal thrust (N) & $F_T$ & Complex amplitude & $\hat{\phi}$ \\
Motor position (m) & $x_M$ & Non-dimensional complex amplitude & $\bar{\phi}$ \\
\end{tabular*}
\end{table}

\section{Model formulation} \label{sec:model_formulation}

\begin{figure}
  \centering
  \includegraphics[width=\textwidth]{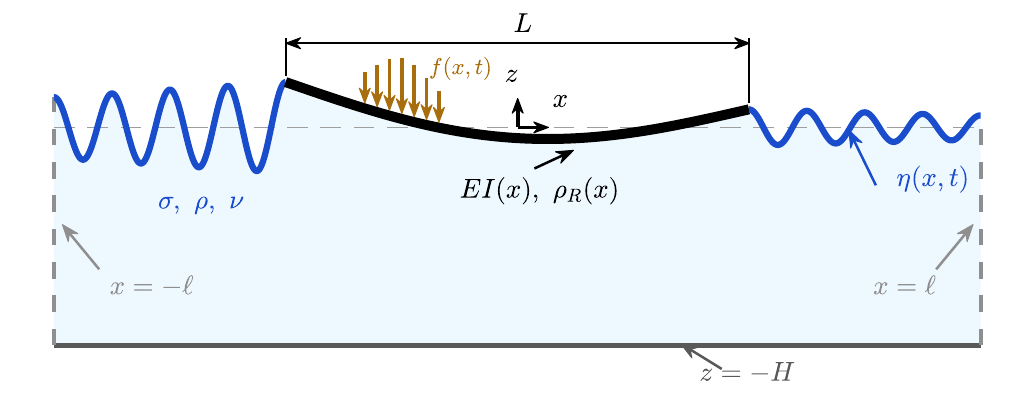}
  
  \vspace{-0.9em}\caption{Schematic of the wave-driven flexible raft model. A thin Euler--Bernoulli raft of length $L$, flexural rigidity $EI(x)$, and mass per unit streamwise length $\rho_R(x)$ lies on the free surface of a weakly viscous fluid of depth $H$, density $
  \rho$, kinematic viscosity $\nu$, and surface tension $\sigma$. The raft is driven by a localized vertical load $f(x,t)$, producing asymmetric gravity--capillary waves that radiate away from the raft in the truncated computational domain
  $x\in[-\ell,\ell]$.}
  \label{fig:schematic}
\end{figure}

We consider the two-dimensional motion of a fluid of depth $H$ with a free surface at $z=\eta(x,t)$ and a thin flexible raft lying along the interface for $-L/2\leq x\leq L/2$ (see figure \ref{fig:schematic}). We work in a rectangular domain in Cartesian coordinates with $z$ measured upward from the undisturbed free surface $z=0$ and with the bottom boundary at $z=-H$. The fluid has constant density $\rho$, kinematic viscosity $\nu$, and surface tension $\sigma$. The raft is modeled as an Euler--Bernoulli beam with flexural rigidity $EI(x)$, mass per unit streamwise length $\rho_R(x)$, out-of-plane width $d$, and is driven by a prescribed vertical load $f(x,t)$ (per unit streamwise length) applied from above. The raft is held against horizontal translation, so the forces computed throughout are predictions for a tethered body.  As such, other effects that might be associated with a free swimmer (such as Doppler shift) are not captured in this model.  However, the tethered assumption circumvents the need for assuming a drag law on the raft, which in general is complicated by the proximity of the interface \citep{hunt2023drag,HarrisBarotta2025}.

Throughout, we assume small-amplitude deformations so that the free-surface and beam equations can be linearised. The raft's equilibrium position accounts for its own weight which rests below the otherwise undisturbed free surface. This static equilibrium solution is time-independent, so a general solution can be decomposed into its static and time-dependent components, $\eta_{s}(x) + \eta(x, t)$, and the linearised equations separate exactly. Solving for the time-independent component is a known, static problem in the context of elastocapillarity \citep{RivettiAntkowiak2013}, and we focus exclusively on the wave-generating time-dependent component in this work.

\subsection{Fluid formulation}

We follow the weakly-viscous quasi-potential free-surface fluid model of \citet{DiasDyachenkoZakharov2008}, which has since been faithfully applied and validated for a number of capillary-scale fluid-structure interaction problems at the interface \citep{galeano2017non, galeano2021capillary, OzaEtAl2023} although fully inviscid theory is capable of predicting the local fluid forces and dynamics of a floating body, even in the presence of weak viscosity \citep{oza2026vertical}. As described previously, \citet{BenhamDevauchelleThomson2024} coupled this fluid model to a thin rigid floating body at the free surface, whose motion reduces to heave, surge, and pitch. We adopt their fluid treatment and momentum-flux accounting of thrust, but now let the raft deform, such that its instantaneous shape is solved together with the flow.  Furthermore, we model the effect of surface tension directly on the body dynamics via a line tension acting at the edges of the raft, where we assume the contact line remains pinned.  

We model the weakly damped flow using a velocity potential $\phi(x,z,t)$ such that $\vect{u}=\nabla\phi \in \mathbb{R}^2$. On the fixed (linearised) domain $-H\le z\le 0$, incompressibility implies
\begin{equation}
\nabla^2\phi = 0, \qquad -H \le z \le 0.
\end{equation}
On $z=0$, the linearised kinematic condition is
\begin{equation}
\phi_z = \mathcal{K}\eta := \eta_t - 2\nu\,\eta_{xx}, \qquad \text{on } z=0.
\label{eqn:kinematic_dim}
\end{equation}
The $2\nu\,\partial_{xx}$ term is the leading viscous correction to the kinematic condition; throughout we retain terms to first order in $\nu$ and neglect $O(\nu^2)$, following the assumptions of \citet{DiasDyachenkoZakharov2008}. The linearised Bernoulli relation at the free surface yields
\begin{equation}
\phi_t + g\eta + \frac{1}{\rho}p + 2\nu\,\phi_{zz} = 0, \qquad \text{on } z=0.
\label{eqn:bernoulli_dim}
\end{equation}
Outside the raft region ($|x|>L/2$), linearized free-surface curvature leads to a normal stress jump from capillarity:
\begin{equation}
p = -\sigma\,\eta_{xx}, \qquad \text{on } z=0,\ |x|>L/2,
\label{eqn:dynamic_harm}
\end{equation}
where $p(x, z, t)$ is the fluid pressure relative to the atmosphere. The bottom boundary is impermeable:
\begin{equation}
\phi_z = 0, \qquad \text{on } z=-H.
\end{equation}
On the free surface, substituting $p=-\sigma\eta_{xx}$ into Bernoulli's equation gives
\begin{equation}
\phi_t + g\eta - \frac{\sigma}{\rho}\eta_{xx} + 2\nu\,\phi_{zz}
=0, \qquad \text{on } z=0,\ |x|>L/2.
\end{equation}
Applying $\mathcal{K}$ to this equation yields
\begin{equation}
\mathcal{K}\!\left(\phi_t + 2\nu\,\phi_{zz}\right) + g\,\phi_z - \frac{\sigma}{\rho}\,\phi_{zxx}
= 0, \qquad \text{on } z=0,\ |x|>L/2,
\end{equation}
which is the corresponding dynamic boundary condition away from the raft.

\subsection{Beam dynamics and bending moment}

The raft's vertical displacement coincides with the interface elevation on $|x|\le L/2$. For a two-dimensional raft of out-of-plane width $d$, the net vertical load per unit streamwise length is $p\,d-f(x,t)$, where $p(x,t) = p(x, z=0, t)$ is the pressure due to the fluid response at the interface and $f(x,t)$ is an applied forcing. Using Bernoulli's equation \eqref{eqn:bernoulli_dim} to eliminate $p$ on the wetted segment gives the forced Euler-Bernoulli beam equation
\begin{equation}
\rho_R(x)\,\eta_{tt} + M_{xx}
= -d\,\rho\left(\phi_t + g\eta + 2\nu\,\phi_{zz}\right) - f(x,t),
\qquad \text{on } z=0,\ |x|\le L/2,
\label{eqn:beam_dim}
\end{equation}
where we note that the constant gravity force term $\rho_R g$ does not appear in \eqref{eqn:beam_dim}, as it contributes to the static equilibrium solution only. Furthermore, we neglect the influence of internal tension of the beam due to surface tension at the end points ($T_{\sigma}\sim\sigma \, d$), therein assuming the beam deformations are dominated by bending resistance (i.e. $ E I/L^2\gg \sigma \, d$, or in the non-dimensional parameters introduced in Section \ref{sec:non-dimensional-form}, $\kappa \gg \Lambda / We $). The bending moment makes the resultant PDE system second order, and relates to the local beam shape $\eta(x,t)$ via
\begin{equation}
M \defeq EI(x)\,\eta_{xx}.
\end{equation}
Applying $\mathcal{K}=\partial_t-2\nu\,\partial_{xx}$ to $M/EI(x)=\eta_{xx}$ and using \eqref{eqn:kinematic_dim} gives
\begin{equation}
\mathcal{K}\!\left(\frac{M}{EI(x)}\right) = \phi_{zxx}, \qquad \text{on } z=0,\ |x|\le L/2,
\end{equation}
allowing the fully coupled problem to be written exclusively in terms of $\phi$ and $M$. This bending-moment formulation allows $EI$ to appear in the denominator, which makes the rigid limit $EI\equiv\infty$ numerically stable to solve. 

At each raft edge the contact line is assumed to be pinned, such that the beam and the free surface meet at the same vertical displacement. Thus the interface elevation is continuous across the transition from raft-covered to uncovered surface:
\begin{subequations}\label{eqn:boundary_conditions_dim}
\begin{align}
\lim_{x \to L/2^{+}} \eta(x,t) &= \lim_{x \to L/2^{-}} \eta(x,t), \\
\lim_{x \to -L/2^{+}} \eta(x,t) &= \lim_{x \to -L/2^{-}} \eta(x,t).
\end{align}
The raft ends are free of externally applied bending moment, so \(M\) vanishes at both endpoints.
\begin{equation}\label{eq:moment_zero}
M(\pm L/2,t) = 0.
\end{equation}
The beam shear force at each endpoint is balanced by the vertical component of the surface-tension traction exerted by the local slope of the free surface \citep{RivettiAntkowiak2013} such that
\begin{align}\label{eq:shear_bc_dim}
\lim_{x\to L/2^-} M_x(x,t) &=\phantom{-}\sigma\,d\,\lim_{x\to L/2^+} \eta_x(x,t), \\
\lim_{x\to -L/2^+} M_x(x,t) &= -\sigma\,d\,\lim_{x\to -L/2^-} \eta_x(x,t),
\end{align}
\end{subequations}
where the signs reflect the opposite orientations of the right and left raft edges.  Note that for the case of $\sigma=0$, the beam boundary conditions on the moment correspond to the traditional free-free boundary conditions.  Equivalently, using \(\mathcal{K}\eta=\phi_z\) on \(z=0\), these end constraints can be written in the variables $(\phi,M)$ as
\begin{subequations}\label{eqn:boundary_conditions_dim_primal}
\begin{align}
\lim_{x \to \pm L/2^{+}} \mathcal{K}^{-1}\phi_z(x,0,t) &= \lim_{x \to \pm L/2^{-}} \mathcal{K}^{-1}\phi_z(x,0,t), & \ \\
M(\pm L/2,t) &= 0, & \ \\
\mathcal{K}M_x(\pm L/2,t) &= \pm \sigma\,d\,\phi_{zx}(\pm L/2^{\pm},0,t).
\end{align}
\end{subequations}
Here $\phi_{zx}(x^{+}, 0, t) = \lim_{y \downarrow x} \phi_{zx}(y, 0, t)$, that is, the limit is taken from above (below if the superscript is negative).

\subsection{Time-harmonic formulation}

Motivated by the application of interest, we study a time-harmonic forcing $f(x,t)$ with angular frequency $\omega$. For a linear system with time-independent coefficients, this forcing produces a response at the same frequency in steady state. Thus, each variable of interest can be written as the real part of a complex time-independent counterpart:
\begin{align*}
\eta(x,t) &= \Real\{\hat{\eta}(x)\mathrm{e}^{i\omega t}\}, &
\phi(x,z,t) &= \Real\{\hat{\phi}(x,z)\mathrm{e}^{i\omega t}\},\\
p(x,t) &= \Real\{\hat{p}(x)\mathrm{e}^{i\omega t}\}, &
M(x,t) &= \Real\{\hat{M}(x)\mathrm{e}^{i\omega t}\},\\
f(x,t) &= \Real\{\hat{f}(x)\mathrm{e}^{i\omega t}\}.
\end{align*}
To distinguish time and frequency representations, we retain the time-domain operator $\mathcal{K}=\partial_t-2\nu\,\partial_{xx}$ and define its harmonic counterpart
\begin{equation}
\hat{\mathcal{K}} \defeq i\omega-2\nu\,\partial_{xx}.
\end{equation}
From the harmonic form of \eqref{eqn:kinematic_dim}, $\hat{\eta}=\hat{\mathcal{K}}^{-1}\hat{\phi}_z$. Substituting this into the harmonic beam equation from \eqref{eqn:beam_dim} and using \eqref{eqn:bernoulli_dim} gives
\begin{equation}
    \hat{M}_{xx} + \hat{\mathcal{L}}\hat{\phi} + \hat{f}(x) = 0,
\end{equation}
where we define
\begin{equation}
\hat{\mathcal{L}}\hat{\phi} \defeq d\rho\left(i\omega \hat{\phi} + 2\nu \hat{\phi}_{zz}\right)
+ (d\rho g - \rho_R(x)\omega^2)\hat{\mathcal{K}}^{-1}\hat{\phi}_{z}.
\end{equation}
In the free surface region ($|x|>L/2$), $\hat{\eta}$ may be eliminated to obtain a single boundary condition for the velocity potential. Combining the harmonic form of \eqref{eqn:bernoulli_dim} with the capillary pressure condition \eqref{eqn:dynamic_harm} gives
\begin{equation}
i\omega \hat{\phi} + g\hat{\eta} - \frac{\sigma}{\rho}\hat{\eta}_{xx} + 2\nu \hat{\phi}_{zz} = 0,
\qquad \text{on } z=0,\ |x|>L/2.
\label{eqn:free-surface-bernoulli}
\end{equation}
Applying \(\hat{\mathcal{K}}=i\omega-2\nu\partial_{xx}\) to \eqref{eqn:free-surface-bernoulli} and using the harmonic form of \eqref{eqn:kinematic_dim} to eliminate \(\hat{\eta}\) gives
\begin{equation}
g\,\hat{\phi}_z - \frac{\sigma}{\rho}\hat{\phi}_{zxx}
= \omega^2 \hat{\phi} - 2i\nu\omega \hat{\phi}_{zz} + 2i\nu\omega \hat{\phi}_{xx} + 4\nu^2 \hat{\phi}_{zzxx}.
\end{equation}
Using Laplace's equation to replace $\hat{\phi}_{zz}=-\hat{\phi}_{xx}$ and neglecting the $O(\nu^2)$ term gives \eqref{eqn:phi-only-free-surface}.
The closed harmonic system for $(\hat{\phi},\hat{M})$ is
\begin{subequations}\label{eqn:governing_harmonic}
\begin{alignat}{2}
\nabla^{2}\hat{\phi} &= 0,
&\quad& \on{-H < z < 0}, \label{eqn:laplace_harm}\\
\hat{\phi}_{z} - \frac{\omega^2}{g}\hat{\phi} - \frac{\sigma}{\rho g}\hat{\phi}_{zxx} - \frac{4i\nu\omega}{g}\hat{\phi}_{xx} &= 0,
&\quad& \on{z=0,\ |x|>L/2}, \label{eqn:phi-only-free-surface}\\
\hat{M}_{xx} + \hat{\mathcal{L}}\hat{\phi} + \hat{f}(x) &= 0,
&\quad& \on{z=0,\ |x|\le L/2}, \label{eqn:beam_phi_m_harm}\\
\hat{\mathcal{K}}\left(\hat{M}/EI(x)\right) &= \hat{\phi}_{zxx},
&\quad& \on{z=0,\ |x|\le L/2}, \label{eqn:moment_phi_m_harm}\\
\hat{M} &= 0,
&\quad& \on{z=0,\ x=\pm L/2}, \label{eqn:moment-edge-harm}\\
\hat{\mathcal{K}}\hat{M}_{x}(\pm L/2) &= \pm d\,\sigma\,\hat{\phi}_{zx}(\pm L/2^{\pm},0),
&\quad& \label{eqn:shear-edge-harm}\\
\hat{\phi}_{z} &= 0,
&\quad& \on{z=-H}. \label{eqn:bottom_harm}
\end{alignat}
\end{subequations}
The raft width $d$ appears as a multiplicative prefactor in the shear force boundary condition and in the fluid forcing terms contained within $\hat{\mathcal{L}}$. All variables and parameters with their corresponding definitions and dimensions are summarized in Table \ref{tab:notation}.

The elevation continuity conditions at the raft edges are imposed in the primal variables as
\begin{align}\label{eqn:boundary_conditions_harm}
\lim_{x \to \pm L/2^{+}} \hat{\mathcal{K}}^{-1}\hat{\phi}_{z}(x,0) &= \lim_{x \to \pm L/2^{-}} \hat{\mathcal{K}}^{-1}\hat{\phi}_{z}(x,0),
\end{align}
When needed for post-processing, the interface elevation and pressure are reconstructed via
\begin{subequations}\label{eqn:reconstruction_harm}
\begin{align}
\hat{\eta} &= \hat{\mathcal{K}}^{-1}\hat{\phi}_{z}, \quad & z = 0, \label{eqn:kinematic_harm} \\
\hat{p} &= -\rho\left(i\omega \hat{\phi} + g\hat{\eta} + 2\nu \hat{\phi}_{zz}\right), \quad & z = 0. \label{eqn:bernoulli_harm}
\end{align}
\end{subequations}

The system \eqref{eqn:governing_harmonic} is solved subject to
\eqref{eqn:boundary_conditions_harm}. The numerical radiation conditions used for horizontal domain truncation are described in Sec.~\ref{sec:numerics}.

This general formulation allows spatially varying $EI(x)$ and $\rho_R(x)$, but the sections below analyze rafts with constant flexural rigidity and mass density. The implementation is available at \texttt{https://github.com/harrislab-brown/flexible\_surferbot}. The code retains the general formulation, and a video of a simulation with non-constant material properties is available as supplementary Movie 1.

\subsection{Non-dimensional form}\label{sec:non-dimensional-form}

For the notation used here, unhatted variables denote dimensional real fields, hatted variables denote dimensional complex Fourier amplitudes, and overbarred variables denote non-dimensional Fourier amplitudes. For the purpose of non-dimensionalizing the equations and identifying dimensionless groups, we take the total beam mass $\rho_R L$ as the characteristic mass, $L$ as the characteristic length, and $\omega^{-1}$ as the characteristic time. The corresponding velocity, potential, pressure, and vertical-load scales are $L\omega$, $L^2\omega$, $\rho_R\omega^2$, and $\rho_R L\omega^2$, respectively. This yields six dimensionless groups that govern the problem
\begin{align}
\label{eq:adim_definition}
\Gamma &= \frac{\rho L^2}{\rho_R}, 
& \quad Fr &= \sqrt{\frac{L \omega^2}{g}}, 
& \quad Re &= \frac{L^2 \omega}{\nu}, \\
\kappa &= \frac{EI}{\rho_R L^4 \omega^2}, 
& \quad We &= \frac{\rho_R L \omega^2}{\sigma},
& \quad \Lambda &= \frac{d}{L}. 
\end{align}
The non-dimensional parameter \(\Gamma\) is a fluid-to-beam mass ratio, whose combination \(\Lambda\Gamma\) gives the ratio of fluid mass below the raft to the beam mass and suggests the relevance of added mass in the problem. \(Fr\) is a Froude number comparing the imposed oscillatory acceleration \(L\omega^2\) with gravity, while \(Re\) is a Reynolds number comparing oscillatory inertia with viscous diffusion. \(\kappa\) is a flexural stiffness parameter comparing elastic bending stiffness with raft inertia at the forcing frequency, such that large \(\kappa\) recovers the rigid limit and smaller \(\kappa\) represents a more flexible raft. \(We\) is a Weber number comparing the raft inertial load scale with surface tension, and \(\Lambda\) is the raft top-view aspect ratio.  As our study is aimed at revealing the phenomenological influence of body stiffness on wave-driven propulsion for the first time, \(\kappa\) is the primary variable of interest for our presented results and subsequent discussion.  

Applying these scalings to \eqref{eqn:governing_harmonic} yields the non-dimensional problem for the complex amplitudes $(\bar{\phi},\bar{M})$. Here and below, $x$ and $z$ denote nondimensional coordinates scaled by $L$ unless stated otherwise. Defining $\bar{\mathcal{K}}=i-\tfrac{2}{Re}\partial_{xx}=\hat{\mathcal{K}}/\omega$, with inverse $\bar{\mathcal{K}}^{-1}=-i-\tfrac{2}{Re}\partial_{xx}=\hat{\mathcal{K}}^{-1}\omega$ to first order in $\nu$, and the linear operator
\begin{equation}
\bar{\mathcal{L}}\bar{\phi} \defeq \Lambda\Gamma\left(i\bar{\phi} - \frac{2}{Re}\bar{\phi}_{xx}\right)
+ \left(\frac{\Lambda\Gamma}{Fr^2}-1\right)\bar{\mathcal{K}}^{-1}\bar{\phi}_{z}=\frac{L}{\rho_R \omega}\hat{\mathcal{L}}\hat{\phi},
\end{equation}
we obtain the non-dimensional governing system and boundary conditions:
\begin{subequations}\label{eqn:nd_final}
\begin{align}
\nabla^{2}\bar{\phi} &= 0, \quad \on{-H/L < z < 0}, \label{eqn:nd_final_a}\\
\bar{\phi} + \frac{4i}{Re}\,\bar{\phi}_{xx} - \frac{1}{Fr^2}\bar{\phi}_{z} + \frac{1}{\Gamma\,We}\,\bar{\phi}_{zxx}
&= 0, \quad \on{z=0,\ |x|>1/2}, \label{eqn:nd_final_b}\\
\bar{M}_{xx} + \bar{\mathcal{L}}\bar{\phi} + \bar{f}(x)
&= 0, \quad \on{z=0,\ |x|\le 1/2}, \label{eqn:nd_final_c}\\
\bar{\mathcal{K}}\left(\bar{M}/\kappa \right) &= \bar{\phi}_{zxx}, \quad \on{z=0,\ |x|\le 1/2}, \label{eqn:nd_final_d}\\
\bar{M} &= 0, \quad \on{z=0,\ x=\pm 1/2}, \label{eqn:nd_final_e}\\
\bar{\mathcal{K}}\bar{M}_x(\pm1/2) &= \pm\frac{\Lambda}{We}\,\bar{\phi}_{zx}(\pm1/2^{\pm},0), \label{eqn:nd_final_f}\\
\bar{\phi}_{z} &= 0, \quad \on{z=-H/L,\ \forall x}. \label{eqn:nd_final_g}
\end{align}
\end{subequations}
The moment relation and the shear condition in \eqref{eqn:nd_final} are written with $\bar{\mathcal{K}}$ on the left because the approximate inverse $\bar{\mathcal{K}}^{-1}=-i-\tfrac{2}{Re}\partial_{xx}$ applied to $\bar{\phi}_{zxx}$ would introduce the fifth-order term $\bar{\phi}_{zxxxx}$.
The continuity conditions in the same primal variables are
\begin{equation}\label{eqn:boundary_conditions_nd}
\lim_{x \to \pm 1/2^{+}} \bar{\mathcal{K}}^{-1}\bar{\phi}_{z}(x,0) = \lim_{x \to \pm 1/2^{-}} \bar{\mathcal{K}}^{-1}\bar{\phi}_{z}(x,0),
\end{equation}

\subsection{Numerical implementation}\label{sec:numerics}

For numerical computations the physical domain is truncated to $x\in[-\ell,\ell]$, and a resulting radiation condition is then imposed. In dimensional variables, outgoing waves satisfy the Sommerfeld condition
\begin{equation}
\hat{\phi}_x = \pm i k\,\hat{\phi}\qquad \text{at } x=\mp \ell
\end{equation}
as in \citet{BenhamDevauchelleThomson2024}.
This closure is appropriate because, sufficiently far from the raft, the radiated wave field is dominated by outgoing traveling modes associated with the complex wavenumber selected by the weakly viscous gravity--capillary dispersion relation,
\begin{equation}
\omega^2 = k\tanh(kH)\left(g+\frac{\sigma}{\rho}k^2\right) + 4i\nu\omega k^2. \label{eqn:dispersion}
\end{equation}
In non-dimensional variables the radiation condition reads
\begin{equation}\label{eqn:radiation_nd}
\bar{\phi}_x = \pm i \bar{k}\,\bar{\phi}\qquad \text{at } x=\mp \bar{\ell},
\end{equation}
with $\bar{k}=kL$ and $\bar{\ell}=\ell/L$. The finite-difference solver is based on the non-dimensional system \eqref{eqn:nd_final}, supplemented with the radiation condition \eqref{eqn:radiation_nd}. We solve this reduced frequency-domain system on a truncated rectangular domain using a tensor-product grid
\[
x_j\in[-\bar{\ell},\bar{\ell}],\quad j=1,\dots,N_x,\qquad
z_m\in[-H/L,0],\quad m=1,\dots,N_z.
\]
The primary algebraic unknowns are the complex potential $\bar{\phi}$ and its $z$-derivative $\bar{\phi}_z$ on all grid points, and the beam moment $\bar{M}$ at nodes on the top boundary where $|x|\le 1/2$. Thus, the total number of unknowns is $N_x N_z + N_x N_z + N_c$, where $N_c$ is the number of nodes on the top boundary where $|x|\le 1/2$.

The bulk operators are at most second order stencils in each unknown. The shear boundary condition \eqref{eqn:nd_final_f} is the sole exception, applying a third $x$-derivative of $\bar{M}$ at the two raft endpoints through a one-sided stencil on those nodes.
Spatial derivatives are built with high-order non-compact finite differences of default formal order $4$, using centered stencils in the interior and one-sided closures near boundaries with the same target order. The 2D derivative matrices are vectorized sparse implementations from a Matlab File Exchange Library \citep{chenyang2026finite}, ported to an open-source Julia implementation. 

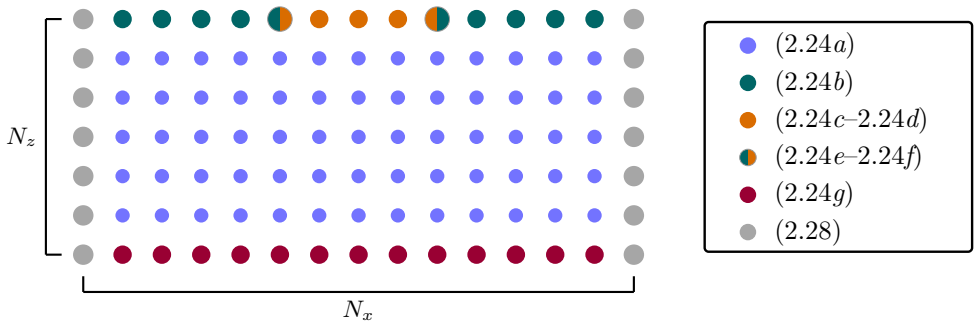
\begin{figure}
  \centering
  \begin{tikzpicture}[x=0.52cm,y=0.52cm]
    \def\Nx{14}
    \def\Ny{6}
    \def\xL{5}
    \def\xR{9}

    \foreach \i in {0,...,14}{
      \foreach \j in {0,...,6}{
        \fill[gray!35] (\i,\j) circle (0.13);
      }
    }

    \foreach \i in {1,...,13}{
      \foreach \j in {1,...,5}{
        \fill[blue!55] (\i,\j) circle (0.182);
      }
    }

    \foreach \i in {0,...,4}{
      \fill[teal!80!black] (\i,\Ny) circle (0.234);
    }
    \foreach \i in {10,...,14}{
      \fill[teal!80!black] (\i,\Ny) circle (0.234);
    }

    \foreach \i in {5,...,9}{
      \fill[orange!85!black] (\i,\Ny) circle (0.234);
    }

    \begin{scope}
      \clip (\xL,\Ny) circle (0.312);
      \fill[teal!80!black] (\xL-0.312,\Ny-0.312) rectangle (\xL,\Ny+0.312);
      \fill[orange!85!black] (\xL,\Ny-0.312) rectangle (\xL+0.312,\Ny+0.312);
    \end{scope}
    \draw[black!40,line width=0.2pt] (\xL,\Ny) circle (0.312);
    \begin{scope}
      \clip (\xR,\Ny) circle (0.312);
      \fill[orange!85!black] (\xR-0.312,\Ny-0.312) rectangle (\xR,\Ny+0.312);
      \fill[teal!80!black] (\xR,\Ny-0.312) rectangle (\xR+0.312,\Ny+0.312);
    \end{scope}
    \draw[black!40,line width=0.2pt] (\xR,\Ny) circle (0.312);

    \foreach \i in {0,...,14}{
      \fill[purple!80!black] (\i,0) circle (0.234);
    }

    \foreach \j in {0,...,6}{
      \fill[gray!70] (0,\j) circle (0.26);
      \fill[gray!70] (\Nx,\j) circle (0.26);
    }

    \draw[line width=0.8pt] (0,-0.95) -- (0,-0.55);
    \draw[line width=0.8pt] (0,-0.95) -- (\Nx,-0.95);
    \draw[line width=0.8pt] (\Nx,-0.95) -- (\Nx,-0.55);
    \node[below] at (\Nx/2,-0.95) {$N_x$};

    \draw[line width=0.8pt] (-0.95,0) -- (-0.55,0);
    \draw[line width=0.8pt] (-0.95,0) -- (-0.95,\Ny);
    \draw[line width=0.8pt] (-0.95,\Ny) -- (-0.55,\Ny);
    \node[left] at (-0.95,\Ny/2) {$N_z$};

    \def\xLeg{16.9}
    \filldraw[fill=white,draw=black,line width=0.9pt,rounded corners=1.5pt] (15.8,0.075) rectangle (22.6,5.925);

    \fill[blue!55] (\xLeg,5.325) circle (0.21);
    \node[anchor=west,font=\normalsize] at (\xLeg+0.45,5.325) {\eqref{eqn:nd_final_a}};

    \fill[teal!80!black] (\xLeg,4.375) circle (0.21);
    \node[anchor=west,font=\normalsize] at (\xLeg+0.45,4.375) {\eqref{eqn:nd_final_b}};

    \fill[orange!85!black] (\xLeg,3.425) circle (0.21);
    \node[anchor=west,font=\normalsize] at (\xLeg+0.45,3.425) {\eqrefrangecompact{eqn:nd_final_c}{eqn:nd_final_d}};

    \begin{scope}
      \clip (\xLeg,2.475) circle (0.21);
      \fill[teal!80!black] (\xLeg-0.21,2.265) rectangle (\xLeg,2.685);
      \fill[orange!85!black] (\xLeg,2.265) rectangle (\xLeg+0.21,2.685);
    \end{scope}
    \draw[black!40,line width=0.2pt] (\xLeg,2.475) circle (0.21);
    \node[anchor=west,font=\normalsize] at (\xLeg+0.45,2.475) {\eqrefrangecompact{eqn:nd_final_e}{eqn:nd_final_f}};

    \fill[purple!80!black] (\xLeg,1.525) circle (0.21);
    \node[anchor=west,font=\normalsize] at (\xLeg+0.45,1.525) {\eqref{eqn:nd_final_g}};

    \fill[gray!70] (\xLeg,0.575) circle (0.21);
    \node[anchor=west,font=\normalsize] at (\xLeg+0.45,0.575) {\eqref{eqn:radiation_nd}};
  \end{tikzpicture}
  \caption{Finite-difference grid used in the truncated computational domain. Purple points impose \eqref{eqn:nd_final_a} in the interior, green points impose \eqref{eqn:nd_final_b} on the free surface outside the raft, orange points impose \eqref{eqn:nd_final_c}--\eqref{eqn:nd_final_d} on the surface in contact with the raft, split green/orange points impose \eqref{eqn:nd_final_e}--\eqref{eqn:nd_final_f} at the raft edge, red points impose \eqref{eqn:nd_final_g} on the bottom, and gray first/last columns impose \eqref{eqn:radiation_nd} as the Sommerfeld condition. We emphasize that split points are shared between the two domains when designing the finite-difference stencil. }
  \label{fig:fd-grid-map}
\end{figure}

The top boundary is partitioned into free-surface nodes ($|x| \geq 1/2$) and nodes below the raft ($|x|\le 1/2$), with one shared split point at each raft/free-surface interface, which ensures \eqref{eqn:boundary_conditions_nd} is satisfied by construction. Equation \eqref{eqn:nd_final_b} is imposed on free-surface nodes; \eqref{eqn:nd_final_c} and \eqref{eqn:nd_final_d} are imposed on nodes below the raft; and edge conditions \eqref{eqn:nd_final_e}--\eqref{eqn:nd_final_f} are enforced by row replacement at the two raft endpoints, eliminating the nullspace of the system, making it invertible. Bottom impermeability \eqref{eqn:nd_final_g} is enforced on $z=-H/L$, and Sommerfeld radiation \eqref{eqn:radiation_nd} is applied on both lateral boundaries over the full vertical extent. Figure \ref{fig:fd-grid-map} shows what equation is applied in each point of the grid.

The complex non-dimensional wavenumber $\bar{k}=kL$ is computed from the weakly viscous gravity--capillary dispersion relation~\eqref{eqn:dispersion}. The default grid places at least $80$ points per horizontal wavelength. The depth is fixed by $\tanh(\Real(\bar{k})H/L)\geq0.99$, matching the deep-water conditions of the reference experiments, although finite depth requires no change to the formulation or the solver. The vertical grid count is then set by $N_z=\lceil 80\,\Real(\bar{k})\,H/L\rceil$, to accurately resolve the vertical dimension, with dynamics expected to decay approximately exponentially away from the interface, as in potential flow theory.

\subsection{Mean thrust and radiation asymmetry}\label{sec:diagnostics}

The non-dimensional system \eqref{eqn:nd_final} determines the time-periodic wave and raft motion, and from this the cycle-averaged thrust can be directly computed. We evaluate it in two independent ways and validate their equality, as required by momentum conservation. The first sums the local forces acting directly on the raft which include the pressure along the raft base and the endpoint capillary forces, projected in the $x$ direction:
\begin{equation}
F_T
=
\avg{ \int_{-L/2}^{L/2} Q(x,t)\,(\vect{n}(x, t)\!\cdot\!\ex)\,\dd x }
+\sigma d\,\Bigl[\avg{\cos\theta}\Bigr]_{x=-L/2}^{x=L/2}.
\label{eq:FTbar_results}
\end{equation}
Here \(Q=p \, d - f\) is the net vertical load on the raft, $\vect{n}(x, t)$ is the unit outward normal to the raft at time $t$, $\avg{\cdot}$ is the time-average of a single cycle, and $\theta(x, t)$ is the angle the free surface makes with the horizontal line $z = 0$ at $x$. This measure of thrust is local and remains valid in this weakly viscous formulation \citep{BenhamDevauchelleThomson2024}. An operational description of how \eqref{eq:FTbar_results} is evaluated from the solution of \eqref{eqn:nd_final} is given in Appendix~\ref{app:thrust}. 

The second method measures the momentum carried away by the radiated waves, using the inviscid radiation stress \citep{LonguetHigginsStewart1964}. In the inviscid far-field for deep-water capillary gravity waves (derivation summarized in \citet{HarrisBarotta2025}), the fore--aft momentum-flux imbalance per unit width due to the outwardly propagating waves is
\begin{equation}
\Delta S_{xx}
=
\left(\frac{1}{4}\rho g+\frac{3}{4}\sigma k^2\right)
\left(|\hat{\eta}(-\ell)|^2-|\hat{\eta}(\ell)|^2\right),
\label{eq:Sxx_farfield_results}
\end{equation}
where \(\hat{\eta}({\ell})\) and \(\hat{\eta}(-{\ell})\) are the complex free-surface amplitudes at the right and left domain boundaries.  The inclusion of viscosity would lead to an exponential attenuation of the wave amplitude away from the body, which introduces a dependence on the computational domain size and is not accounted for in this expression.  We take positive \(F_T\) as thrust in the \(+x\) direction. A body recoils opposite to the net momentum it radiates, so the raft is propelled toward \(+x\) when it radiates waves more strongly toward \(-x\) (i.e. the wave amplitude is larger at the stern of a wave-driven craft). We therefore write the imbalance in \eqref{eq:Sxx_farfield_results} as the left intensity minus the right, so that \(F_T=d\,\Delta S_{xx}\) holds with no sign change and the horizontal force on the raft equals the net momentum carried off by the radiated waves. This imbalance differs by a sign from the Longuet--Higgins radiation-stress difference, which is the force required to hold the raft fixed rather than the thrust it develops.

The far-field inviscid wave measure \eqref{eq:Sxx_farfield_results} is a positive prefactor times the squared difference of outgoing wave intensities \(|\hat{\eta}(-\ell)|^2-|\hat{\eta}(\ell)|^2\). Normalizing that difference by their sum yields a non-dimensional proxy for the efficiency of wave propulsion represented as an asymmetry factor
\begin{equation}
  \alpha = \frac{|\bar{\eta}(-\bar{\ell})|^{2} - |\bar{\eta}(\bar{\ell})|^{2}}
               {|\bar{\eta}(\bar{\ell})|^{2} + |\bar{\eta}(-\bar{\ell})|^{2}}.
  \label{eq:alpha_def_app}
\end{equation}
Here \(\alpha=0\) corresponds to equal outgoing intensities, and hence no net thrust. The limiting case of \(\alpha=\pm1\) corresponds to radiation on one side only, and thus all radiated waves directly contribute to forward propulsion, interpreted as an efficient use of waves. In particular the numerator scales with the net wave thrust and the denominator with the total wave energy, and can be interpreted as a proxy for the thrust-to-power ratio, a common metric used to characterize and compare propulsion systems.
In this sense \(\alpha\) is distinct from the propulsive efficiency of \citet{BenhamDevauchelleThomson2024} and \citet{HarrisBarotta2025}. They define efficiency as the ratio of the useful propulsive power (thrust times drift speed) to the applied actuator power. However this calculation necessarily requires a drift speed, which is not applicable to the tethered case considered here. As such \(\alpha=\pm1\) does not necessarily imply high propulsion efficiency; it is a necessary but not sufficient condition.  

The difference between the two outgoing intensities can also be written as
\begin{equation}
   |\bar{\eta}(-\bar{\ell})|^2 - |\bar{\eta}(\bar{\ell})|^2 = -4 \,\Real(SA^*) = -4 \,|S|\, |A| \cos\left(\arg(S) - \arg(A)\right).
  \label{eq:SA_decomp_app}
\end{equation}
where the free-surface elevation has been split into symmetric and antisymmetric parts,
\begin{equation}
  S = \frac{\bar{\eta}(\bar{\ell})+\bar{\eta}(-\bar{\ell})}{2},
  \qquad
  A = \frac{\bar{\eta}(\bar{\ell})-\bar{\eta}(-\bar{\ell})}{2},
  \quad\bar{\eta}(\bar{\ell})=S+A,\quad \bar{\eta}(-\bar{\ell})=S-A. 
  \label{eq:SA_def_app}
\end{equation}
Substituting into our definition of $\alpha$ in \eqref{eq:alpha_def_app} gives
\begin{equation}
  \alpha=\frac{-2\,\Real(SA^*)}{|S|^2+|A|^2} , \quad F_T \propto -4\, \Real(SA^*).
  \label{eq:alpha_SA_app}
\end{equation}
This factorization shows that net thrust requires both $S$ and $A$ to be nonzero, with a phase difference other than $\pi/2$ modulo $\pi$. A purely symmetric response ($A=0$) or purely antisymmetric response ($S=0$) radiates equally in both directions. When $S\perp A$, their complex amplitudes differ in phase by $\pi/2$. The symmetric and antisymmetric contributions do not interfere, as they are out-of-phase at the domain boundaries. Adding or subtracting a perpendicular complex amplitude changes the phase of the wave but not its overall magnitude, since $|S+A|^2=|S-A|^2=|S|^2+|A|^2$.  As such, this represents a third case (in addition to $A=0$ or $S=0$) where the waves are of equal amplitude on each side, and produce no net thrust.  This calculation will be useful in interpreting our results.

\subsection{Reduced-order modal model and the parity structure of thrust}\label{sec:modal-reduction}

For a raft with constant material properties, we may reduce the coupled system \eqref{eqn:nd_final} to a model that is significantly less computationally expensive while also facilitating more physical interpretation as to the role of flexibility in the propulsion through modal analysis \citep{Newman1994,korobkinEigenmodesAddedmassMatrices2023}. As was suggested in the prior section and we will further demonstrate here, symmetry plays an important role in the understanding of wave force and power \citep{falnes2015fundamental}. 

To begin the process, we expand the raft displacement, hydrodynamic load and forcing profile in the free--free Euler--Bernoulli modal basis $W_n(x)$ on $x\in[-1/2,1/2]$ (see \citet{rao1995mechanical}, reviewed in Appendix~\ref{app:freefree-modes}):
\begin{equation}
\bar{\eta}(x)=\sum_{n=0}^{N-1}\bar q_nW_n(x),\qquad
\bar p_{\mathrm{dyn}}(x)=\sum_{n=0}^{N-1}\bar p_nW_n(x),\qquad
\bar f(x)=\sum_{n=0}^{N-1}\bar f_nW_n(x).
\label{eq:modal_decomposition}
\end{equation}
Here $\bar p_{\mathrm{dyn}}=-\left(i\bar\phi-2\bar\phi_{xx}/Re\right)$ is the non-dimensional dynamic pressure. In the context of hydroelastic problems, these are sometimes referred to as the ``dry'' modes \citep{korobkinEigenmodesAddedmassMatrices2023}.  An exact expansion has $N=\infty$, while a reduced model retains a finite number of modes. Modes $n=0$ and $n=1$ describe rigid-body translation and rotation modes, respectively, whereas all modes $n\geq2$ describe elastic bending modes. The modes are $L^2$-orthonormal, so the modal coefficients are
\begin{equation}
\bar q_m=\int_{-1/2}^{1/2}\bar\eta\, W_m\,\dd x,\qquad
\bar p_m=\int_{-1/2}^{1/2}\bar p_{\mathrm{dyn}}W_m\,\dd x,\qquad
\bar f_m=\int_{-1/2}^{1/2}\bar f\,W_m\,\dd x.
\label{eq:modal_load_definitions}
\end{equation}
The spectral decomposition and its supporting algebra are given in Appendix~\ref{app:added-mass}. In non-dimensional form, the force balance in the raft \eqref{eqn:beam_dim} becomes
\begin{equation}
\kappa\bar{\eta}_{xxxx}
+\left(\frac{\Lambda\Gamma}{Fr^2}-1\right)\bar{\eta}
=\Lambda\Gamma\,\bar{p}_{\mathrm{dyn}}-\bar{f},
\qquad |x|\le 1/2.
\label{eq:nd_beam_modal_start}
\end{equation}
Here $\kappa\bar\eta_{xxxx}$ is the elastic restoring term, $\Lambda\Gamma\bar\eta/Fr^2$ is the hydrostatic restoring term, and $-\bar\eta$ is the raft inertia.  Projecting \eqref{eq:nd_beam_modal_start} onto $W_m$ gives the modal equation for mode $m$
\begin{equation}
\left(\kappa\beta_m^4+\frac{\Lambda\Gamma}{Fr^2}-1\right)\bar{q}_m
=\Lambda\Gamma\,\bar{p}_m-\bar{f}_m-\frac{\Lambda}{We}\bar{K}_m^\sigma,
\label{eq:modal_balance_adim_app}
\end{equation}
where $\beta_m$ are the free--free eigenvalues and satisfy $W_n''''=\beta_n^4W_n$. The terms on the left are the elastic, hydrostatic, and inertial contributions to mode $m$.  Three additional modal forcing terms appear on the right-hand side of the equation.  The first term is the effective hydrodynamic load on the mode. The second term measures the projected external forcing onto the mode (modal forcing), which depends on the  amplitude and spatial structure of the forcing (the strength and location of the motor, for instance).  The final quantity $\bar K_m^\sigma$ is the capillary boundary term
\begin{equation}
\bar K_m^\sigma
=W_m(1/2)\bar\eta_x(1/2^+)
 +W_m(-1/2)\bar\eta_x(-1/2^-),
\label{eq:modal_capillary_definition}
\end{equation}
which accounts for the capillary force at the ends of the raft. 

By linearity of the full problem \eqref{eqn:nd_final}, the hydrodynamic and capillary contributions are themselves linear maps of the modal amplitudes. Thus, we introduce the hydrodynamic impedance $\bar{\mathbf Z}$ and capillary endpoint $\bar{\mathbf C}^{\sigma}$ matrices,
\begin{equation}
\bar p_m=\sum_{n=0}^{N-1}\bar Z_{mn}\bar q_n,
\qquad
\bar K_m^\sigma=\sum_{n=0}^{N-1}\bar C_{mn}^\sigma\bar q_n.
\label{eq:modal_linear_maps}
\end{equation}
Here $\bar{\mathbf Z}$ maps a prescribed raft shape to its dynamic-pressure load and $\bar{\mathbf C}^\sigma$ maps it to the two capillary edge forces. Define
\begin{equation}
\bar D_{mn}(\kappa)
=\left(\kappa\beta_m^4+\frac{\Lambda\Gamma}{Fr^2}-1\right)\delta_{mn},
\qquad
\bar{\mathbf M}(\kappa)
=\bar{\mathbf D}(\kappa)-\Lambda \left(\Gamma\bar{\mathbf Z}
 -\frac{1}{We}\bar{\mathbf C}^\sigma \right).
\label{eq:modal_matrix_definition}
\end{equation}
Substituting these maps into \eqref{eq:modal_balance_adim_app} gives the reduced-order modal model
\begin{equation}
\bar{\mathbf M}(\kappa)\,\bar{\boldsymbol q}
=-\bar{\boldsymbol{f}}.
\label{eq:modal_radiation_map}
\end{equation}

Here $\bar{\mathbf M}(\kappa)$ is the complex modal impedance, which includes the effects of bending, raft inertia, hydrostatic pressure, dynamic pressure, and capillarity.  The forcing term $\bar{\boldsymbol{f}}$ is the modal forcing vector, which depends only on the modal basis and the external forcing.  The matrix  $\bar{\mathbf M}(\kappa)$ depends on all fluid and raft parameters, as well as the frequency of the external forcing.  For fixed geometry, frequency, fluid parameters, radiation boundary conditions, and modal basis $W_n$, $\bar{\mathbf Z}$ and $\bar{\mathbf C}^\sigma$ are not functions of $\Lambda$, $\Gamma$, forcing amplitude, forcing position, or $\kappa$. Those quantities enter only through $\bar{\boldsymbol f}$, $\bar{\mathbf D}$, and the prefactors $\Lambda\Gamma$ and $\Lambda/We$.

Once $\bar{\mathbf Z}$ and $\bar{\mathbf C}^{\sigma}$ (and thus $\bar{\mathbf M}$) are known, the modal amplitudes $\bar{\boldsymbol q}$ can be calculated for any forcing vector $\bar{\boldsymbol f}$, from which the raft dynamics can be fully constructed and thrust computed. We construct these matrices by prescribing one mode at a time, with $\bar\eta=W_n$ on the raft. Each solve for the fluid response \eqref{eqn:nd_final} to the imposed kinematics gives the $n$th columns of $\bar{\mathbf Z}$ and $\bar{\mathbf C}^{\sigma}$, together with the corresponding velocity potential, bending moment distribution, and modal wave coefficient
\begin{equation} \label{eq:radiation_coeff_def}
\bar{a}_n=\bar\eta(\bar\ell)\big|_{\bar\eta=W_n}.
\end{equation}
After $N$ such solves of the full model, one may reconstruct the response to any applied forcing via the linear superposition principle, provided one keeps the geometry, frequency, and fluid parameters fixed. In particular, changing the raft stiffness (via $\kappa$), the shape of the applied forcing or the coupling constant $\Lambda$ requires only the solution of the $N\times N$ linear system \eqref{eq:modal_radiation_map}, facilitating remarkably efficient exploration of the influence of raft flexibility on the coupled fluid-structure problem.

Continuing to assume uniform material properties, the raft and fluid domain have left--right symmetry. As such, in the basis solve wherein we prescribe $\bar\eta=W_n$, the pressure and capillary loads have the same symmetric ($W_n(x) = W_n(-x)$) or antisymmetric ($W_n(x) = -W_n(-x)$) parity as the prescribed mode.  Even numbered modes are spatially symmetric, whereas odd numbered modes are antisymmetric.  As such, any single mode in isolation is incapable of producing thrust, thus necessitating a blend of modes to achieve propulsion via wave interference. It is shown in Appendix~\ref{app:modal-parity} that projections onto any mode of the opposite parity $(i \not\equiv j \text{ (mod } 2)) $ implies  $\mathbf{\bar M}_{i, j} = 0$. We can thus rearrange the rows and columns of $\bar{\mathbf M}$, giving the block-diagonal system \citep{korobkinEigenmodesAddedmassMatrices2023}
\begin{equation}
\bar{\mathbf M}(\kappa)
=\begin{pmatrix}
\bar{\mathbf M}_e(\kappa)&\mathbf 0\\
\mathbf 0&\bar{\mathbf M}_o(\kappa)
\end{pmatrix},
\qquad
\bar{\boldsymbol f}
=\begin{pmatrix}\bar{\boldsymbol f}_e\\\bar{\boldsymbol f}_o\end{pmatrix}.
\label{eq:modal_parity_blocks}
\end{equation}
The even (symmetric) and odd (antisymmetric) modal amplitudes decouple and can therefore be calculated independently. An asymmetric forcing profile generally has non-zero projections onto both blocks, while changing the raft flexibility (via $\kappa$) changes the amplitudes and phases produced within each block.

We now use this parity structure to separate the effects of flexibility and forcing position on the asymmetry factor $\alpha$ and thrust $F_T$. Since the far-field wave amplitudes \(\bar \eta(\pm\bar \ell)\) depend linearly on $\bar{\boldsymbol q}$, their symmetric and antisymmetric parts $S$ and $A$ are also linear in the modal amplitudes \(\mathbf{\bar q}\). Left--right symmetry means that an even mode produces equal elevations at the two boundaries and therefore contributes only to $S$. An odd mode produces boundary elevations of opposite sign and contributes only to $A$. Collecting the radiation coefficients $\bar a_n$ from equation \eqref{eq:radiation_coeff_def} into the even and odd modes $\boldsymbol{\bar a}_e$ and $\boldsymbol{\bar a}_o$, gives
\begin{equation}
S=\boldsymbol{\bar a}_e^{\mathsf T}\bar{\boldsymbol q}_e,
\qquad
A=\boldsymbol{\bar a}_o^{\mathsf T}\bar{\boldsymbol q}_o.
\label{eq:modal_SA_radiation}
\end{equation}
Combining these modal wave coefficients with the solutions of the two modal blocks \eqref{eq:modal_parity_blocks} gives
\begin{align}
\label{eq:modal_transfer_rows}
S&=\boldsymbol{\bar r}_e(\kappa)\bar{\boldsymbol f}_e,
&\boldsymbol{\bar r}_e(\kappa)&=-\boldsymbol{\bar a}_e^{\mathsf T}\bar{\mathbf M}_e(\kappa)^{-1},\\
A&=\boldsymbol{\bar r}_o(\kappa)\bar{\boldsymbol f}_o,
&\boldsymbol{\bar r}_o(\kappa)&=-\boldsymbol{\bar a}_o^{\mathsf T}\bar{\mathbf M}_o(\kappa)^{-1}.
\label{eq:modal_transfer_rows2}
\end{align}
The rows $\boldsymbol{\bar r}_e$ and $\boldsymbol{\bar r}_o$ include both the modal response of the raft and the conversion of that response into outgoing waves. This relationship between the forcing coefficients and the symmetric/antisymmetric components of thrust is general in form. Inspired by experiments, in the present work we restrict our investigation to a one-dimensional family of localized Gaussian forcing profiles with fixed breadth and magnitude, parametrised only by the center of the forcing, $x_M/L$. For all values of $\kappa$, one finds a discrete set of forcing positions $x_M/L$ for which $S = 0$ (or $A = 0$). This happens when the even forcing vector $\bar{\boldsymbol f}_e$  is orthogonal to $\boldsymbol{\bar r}_e(\kappa)$ (equivalently for $\bar{\boldsymbol f}_o$).

Substitution of \eqref{eq:modal_transfer_rows} and \eqref{eq:modal_transfer_rows2} gives
\begin{equation}
F_T, \,\alpha \propto \Real(SA^*)
=\bar{\boldsymbol f}_e(x_M/L)^{\mathsf T}
 \mathbf G(\kappa)
 \bar{\boldsymbol f}_o(x_M/L),
\label{eq:modal_load_family_condition}
\end{equation}
where we have used the fact that the forcing load is a real vector, and we define the real matrix
\begin{equation}
\mathbf G(\kappa)
=\Real\!\left(\boldsymbol{\bar r}_e(\kappa)^{\mathsf T}
              \boldsymbol{\bar r}_o(\kappa)^*\right).
\label{eq:modal_interference_matrix}
\end{equation}
Equation \eqref{eq:modal_load_family_condition} decouples the effects of flexibility and forcing as they relate to propulsive efficiency and thrust. Stiffness changes $\mathbf G$, while motor position changes the projections of the imposed load onto the even and odd modes ($\bar{\boldsymbol f}_e$ and $\bar{\boldsymbol f}_o$). For the inviscid problem, $\mathbf G(\kappa)=\mathbf 0$ at isolated values of $\kappa$. At these stiffnesses, \eqref{eq:modal_load_family_condition} gives $\Real(SA^*)=0$ (no thrust) for any arbitrary forcing profile. In the hydrodynamically uncoupled limit (corresponding to $\Lambda=0$), these discrete $\kappa$ values correspond exactly to resonances of the free-free beam elastic modes (i.e. $\kappa=\beta_n^{-4}$, giving $1.998\times10^{-3}$, $2.629\times10^{-4}$, $6.841\times10^{-5}$, \dots, see also Appendix~\ref{app:uncoupled-reference}), while at the reference width $\Lambda = 0.6$ hydrodynamic coupling raises them by approximately a factor of three.  

Each parity block contains one rigid mode and \(\left\lfloor\frac{N-2}{2}\right\rfloor\) elastic modes. The stiffness condition $\mathbf G(\kappa)=\mathbf 0$ reduces to a real polynomial $\mathcal P(\kappa) = 0$ of degree $N-2$ (Appendix~\ref{app:out-of-phase}). The polynomial is assembled from the same $N$ solves of the full model used to construct \eqref{eq:modal_radiation_map}.

\section{Results}\label{sec:results}

\subsection{Reference rigid-raft response}

\begin{figure}
  \centering
  \captionsetup[subfigure]{aboveskip=2pt, belowskip=0pt}
  \includegraphics[width=\textwidth]{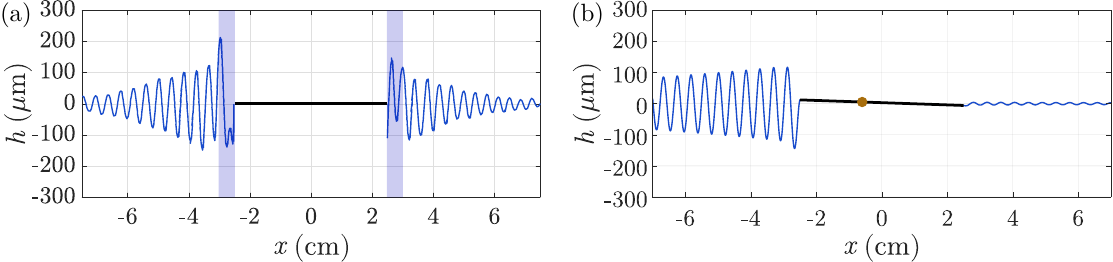}
  \par\vspace{0.1em}
  \begin{subfigure}[t]{0.48\textwidth}
    \centering
    \begin{tikzpicture}
      \node[anchor=south west, inner sep=0] (image) at (0,0) {\includegraphics[width=\textwidth]{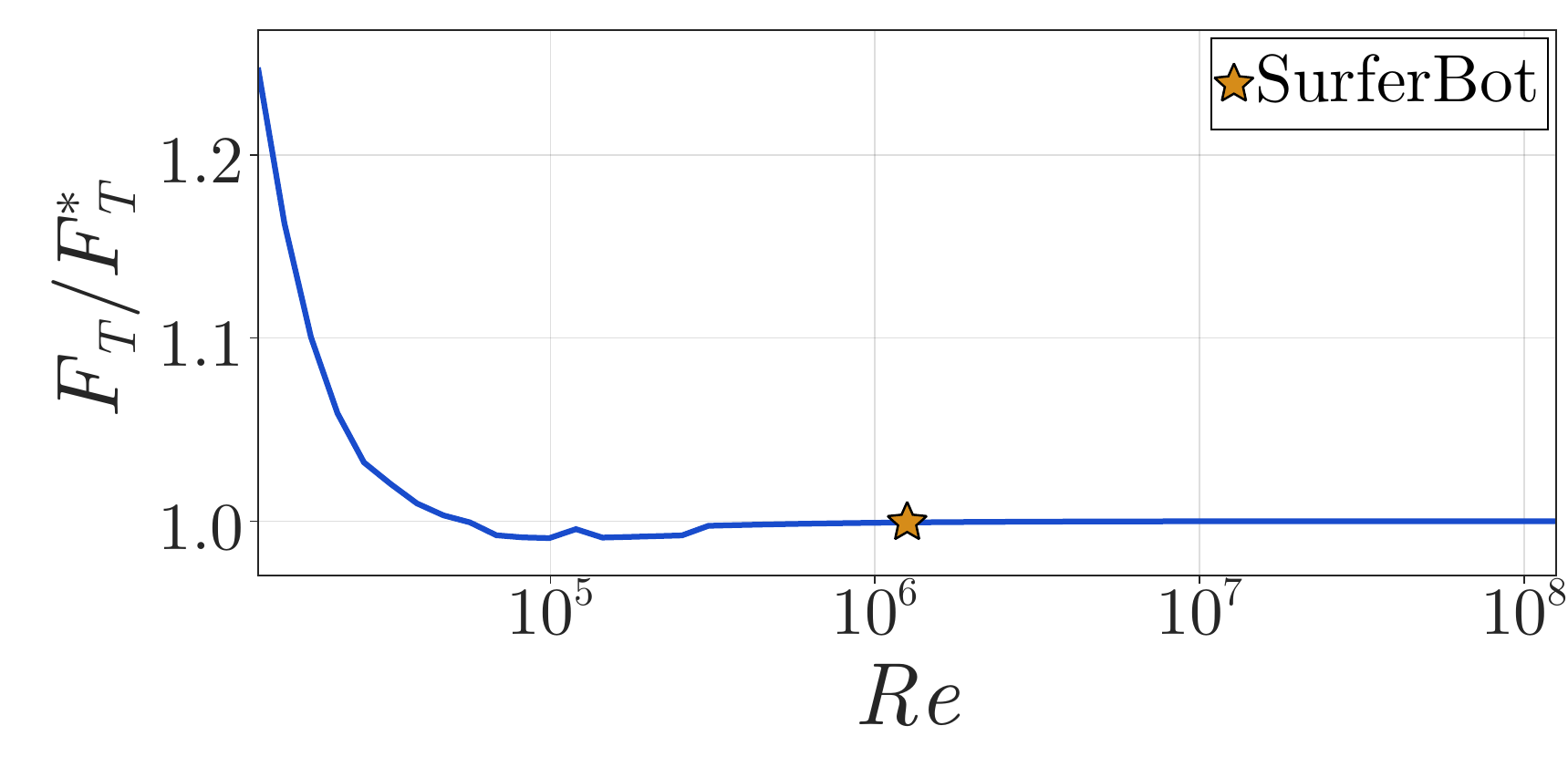}};
      \node[anchor=north west, inner sep=0pt, xshift=0.0264\textwidth, yshift=-8pt] at (image.north west) {\scalebox{0.9588}{(c)}};
    \end{tikzpicture}
  \end{subfigure}
\caption{Rigid limit and predicted dependence on Reynolds number. Panel (a) shows the SurferBot wavefield reported by \citet{RheeEtAl2022}, with a shaded blue region where measurements are not reliable due to limitations of the interface reconstruction method. Panel (b) shows the corresponding prediction for a rigid raft from the present formulation, computed with \(\kappa=\infty\); the dot marks the motor position. Panel (c) shows \(F_T/F_T^\ast\) as a function of \(Re=L^2\omega/\nu\), where \(F_T^\ast\) is the predicted thrust for the rigid Surferbot reference in the inviscid limit, as defined in \eqref{eqn:fstar_def}. The star marks the water/SurferBot operating point.}
  \label{fig:benham_aguero_validation}
\end{figure}

Figure~\ref{fig:benham_aguero_validation} first compares the rigid limit (\(\kappa=\infty\)) of the present formulation (panel b) against the SurferBot wavefield of \citet{RheeEtAl2022} (panel a). Despite being a fully 2D model, the numerical calculation produces the same qualitative asymmetry between the left and right waves as the reference SurferBot response with a comparable peak wave amplitude behind the craft. Faster decay of the waves is observed in the experiment due to the radial spreading not accounted for in the 2D model, as noted by \citet{BenhamDevauchelleThomson2024}. Supplementary Movie 2 animates this rigid-limit response over ten forcing periods.  

To set this reference point, we use water properties with \(\rho=1000\,\mathrm{kg\,m^{-3}}\), \(\sigma=72.2\times10^{-3}\,\mathrm{N\,m^{-1}}\), and \(g=9.81\,\mathrm{m\,s^{-2}}\), together with a raft of length \(L=0.05\,\mathrm{m}\) and width \(d=0.03\,\mathrm{m}\). The SurferBot of \citet{RheeEtAl2022} has a total mass of \(2.6\,\mathrm{g}\), comprising the printed base, the vibration motor and the battery. We distribute this mass uniformly along the raft, giving a constant \(\rho_R=0.052\,\mathrm{kg\,m^{-1}}\).  The motor drives the raft through an eccentric mass \(m_M = 0.13 \) g rotating at radius \(r_M = 2.5\) mm  about the motor axis, which exerts a rotating force of magnitude \(m_M r_M\omega^2\), with eccentric-mass moment \(I_M=m_Mr_M=3.25\times10^{-7}\,\mathrm{kg\,m}\). The imposed load has a Gaussian profile with width \(0.05L\), centered at a position $x_M/L=-0.12$, restricted to $|x|\le L/2$ and renormalized to the target total force. These dimensional quantities define the reference non-dimensional values \(\Gamma=48.1\), \(Fr=35.9\), \(We=9.10\times10^3\), and \(\Lambda=0.6\), as defined in Table \ref{tab:notation}, and are held fixed unless stated otherwise.

A sweep over Reynolds number is presented in figure~\ref{fig:benham_aguero_validation}(c), demonstrating that the normalized thrust is largely insensitive to \(Re\) near the water/SurferBot operating point. Viscosity remains part of the model with weak dissipation, but over the range relevant here it does not influence the propulsion trends. The thrust as given by \eqref{eq:FTbar_results} (Section~\ref{sec:diagnostics}) is determined by pressure on the raft and capillary forces at the raft ends.  While viscosity does influence the far-field decay of the radiated waves \citep{OzaEtAl2023,sun2026}, it does not have a strong influence on the local body forces and dynamics, as also evidenced in the work of \citet{oza2026vertical}.
We therefore set \(\nu=0\) in all subsequent parameter sweeps below since the focus of the present study is on propulsion specifically.  We also define a reference thrust for the purposes of our study as
\begin{equation}
F_T^\ast=F_T\big|_{\nu=0,\,\kappa=\infty},
\label{eqn:fstar_def}
\end{equation}
that is, the thrust evaluated at the inviscid, rigid SurferBot reference point with the reference geometry, forcing, and frequency described above.

As measured by \citet{RheeEtAl2022}, an off-centre motor loads the two sides of the raft unevenly, producing unequal outgoing waves and a non-zero mean thrust. In the rigid limit (\(\kappa=\infty\)), only the two rigid-body modes participate, heave (\(n=0\)) and pitch (\(n=1\)), and the net thrust depends on how the applied force projects onto each. Flexibility adds shape degrees of freedom, so the same motor force is then distributed among heave, pitch, and elastic modes. The remaining results explore how the participation of the elastic modes changes the outgoing wave imbalance and can strengthen, weaken, eliminate, or even reverse the rigid-raft thrust.

\subsection{Influence of motor position and raft flexibility}

\begin{figure}
  \centering
  \includegraphics[width=\textwidth]{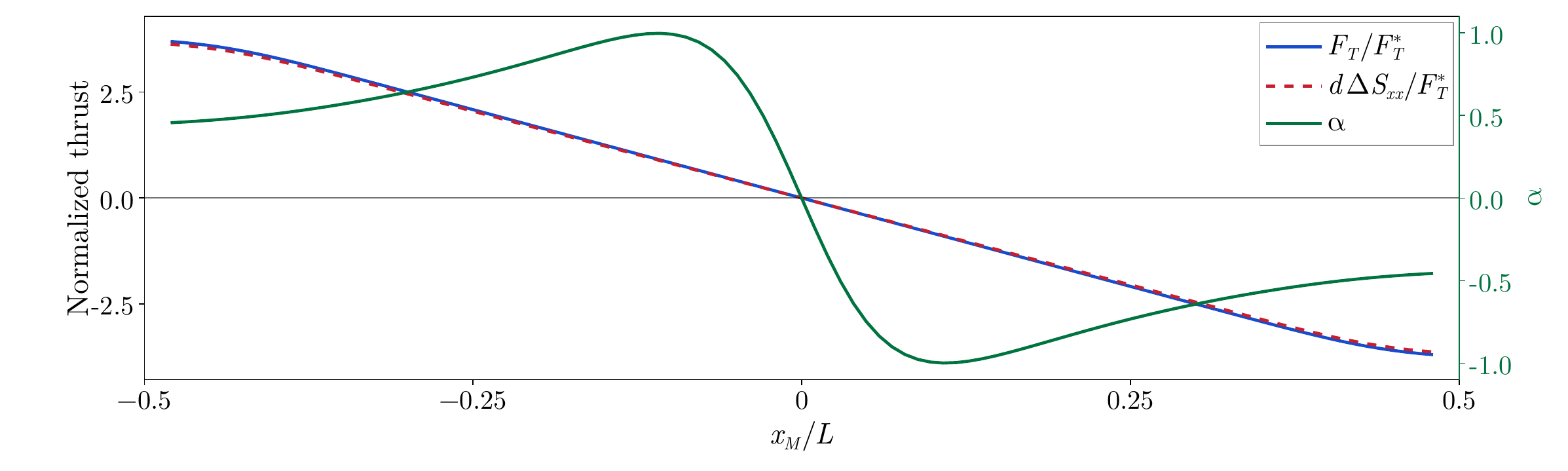}
  \caption{Motor-position sweep for a rigid raft at the SurferBot operating point. The normalized thrust \(F_T/F_T^\ast\) is shown against motor position \(x_M/L\) in the rigid (\(\kappa=\infty\)) and inviscid (\(Re=\infty\)) limits. The blue line gives the direct thrust, the red dashed line the Longuet--Higgins radiation-stress prediction, and the green curve the far-field asymmetry factor \(\alpha\).}
  \label{fig:rigid_position_sweep}
\end{figure} 

To begin our main results, we vary the motor position and the flexural rigidity separately about the SurferBot operating point to explore their influence on the thrust production and wave asymmetry. Figure~\ref{fig:rigid_position_sweep} shows a rigid-limit sweep where only the motor position is varied, in which the thrust grows as the motor moves from the centre and is largest when placed at the raft end ($x_M/L = \pm0.5$). Although not directly measured in the work, \citet{RheeEtAl2022} note that the SurferBot tended to move faster when the motor was positioned closer to the end.  Figure~\ref{fig:rigid_position_sweep} also shows an overlay of the near-field and the far-field thrust measures, $F_T$ and $d\,\Delta S_{xx}$ respectively, which agree to three significant digits in this range of study, and serve as further validation of our numerical method.  While the maximum thrust occurs when the motor is placed at the end, this does not correspond to the most asymmetric wavefield, as captured by the asymmetry factor $\alpha$.  In particular we also observe that the most asymmetric wavefield occurs at $x_M/L \approx \pm0.11$, a location that approaches $\pm1/6$ in the hydrodynamically uncoupled limit ($\Lambda \to 0$), which we derive analytically in Appendix \ref{app:rigid-optimum}. Curiously, this predicted position is close to the experimental conditions of the reference SurferBot, and corresponds to a high (wave) thrust-to-power ratio, as described earlier. Note that figure~\ref{fig:rigid_position_sweep} spans both signs of \(x_M\) and recovers the anticipated anti-symmetry in both measures (\(F_T(-x_M)=-F_T(x_M)\) and \(\alpha(-x_M)=-\alpha(x_M)\)) as expected for the uniform beam case. As such, the remaining sweeps show \(x_M/L\le0\) only.

\begin{figure}
  \centering
  \begin{tikzpicture}
    \node[anchor=south west,inner sep=0] (gridimage) at (0,0) {\includegraphics[width=\textwidth]{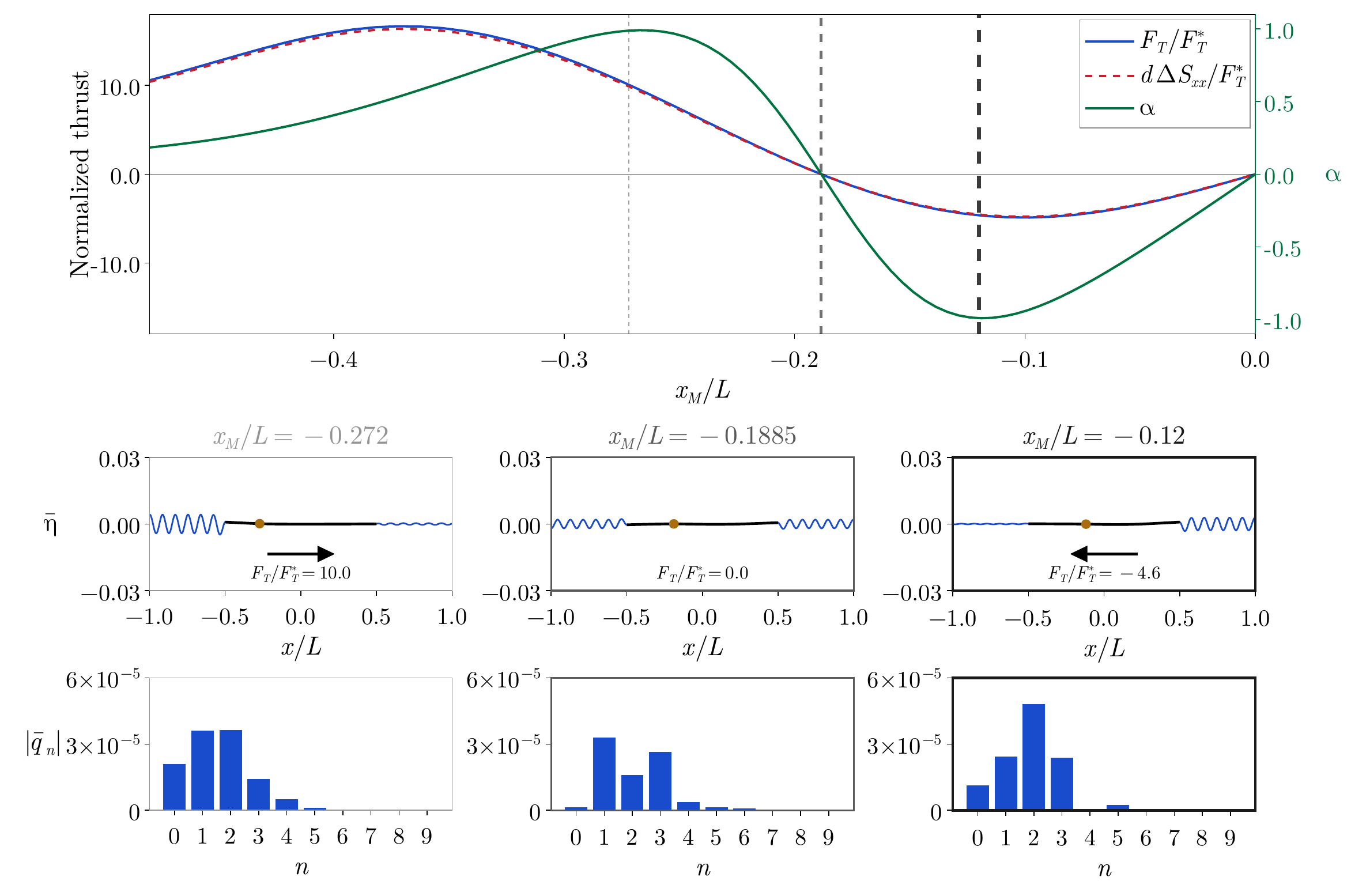}};
    \gridpaneltags
  \end{tikzpicture}
  \caption{Motor-position sweep and representative responses for a flexible raft. The plot in (a) shows the normalized thrust and the asymmetry factor as a function of forcing position \(x_M/L\) at fixed \(\kappa=6.87\times10^{-3}\). The blue line gives the direct thrust, the red dashed line is the Longuet--Higgins radiation-stress prediction, and the green curve is the asymmetry factor \(\alpha\). Vertical dashed lines mark the three motor positions shown in panels (b)--(d) with each color and thickness coded to match the corresponding  panel below. The middle row shows the corresponding free-surface profiles, with the raft contact segment in black and the motor position marked in gold; the arrow and the labeled value below each raft give the sign and magnitude of \(F_T/F_T^\ast\) at that operating point. The bottom row (e)--(g) shows the modal amplitudes \(|\bar{q}_n|\).}
  \label{fig:motor_position_grid}
\end{figure}

Figure~\ref{fig:motor_position_grid} shows the same motor position sweep, but at a finite flexibility of \(\kappa=6.87\times10^{-3}\), a value where we see notable departures from the rigid case.  For context, in the hydrodynamically uncoupled limit, the lowest flexible mode ($n=2$) is resonant at a $\kappa$ value of $1.998\times10^{-3}$.  The original 3D-printed SurferBot has an estimated non-dimensional stiffness of $\kappa \approx 0.3$ that accounts for the rigidifying support structures atop the base, and is thus well approximated as rigid in the present study.  This estimate is determined from a static finite-element analysis conducted in Autodesk Fusion where the SurferBot is pinned along its leading and trailing edges, a normal point load is applied to the center, and the effective stiffness is inferred from the computed maximum displacement.  
The predictions for the flexible beam are qualitatively different from the rigid sweep.  First, unlike the rigid case, the flexible case does not always travel with the motor at the rear of the craft, but can actually reverse direction depending on its exact placement.  Secondly, the motor position associated with peak thrust is no longer at the end, but now at an interior forcing position.  For these parameters, this maximum occurs near \(x_M/L\approx-0.37\).  Lastly, the predicted maximum thrust is significantly enhanced by the introduction of flexibility with a maximum of \(F_T/F_T^\ast\approx16.7\), about $4.5$ times the maximum value in the rigid position sweep.  As before, the locations of maximum wave asymmetry and maximum thrust do not occur at the same position.  Thrust and wave asymmetry always carry the same sign, by definition.

We also include three representative snapshots of the wavefield in panels (b)--(d) of figure~\ref{fig:motor_position_grid}. They similarly showcase how the wave thrust can be directed either way by changing the motor position on one side of the raft, a feature distinct from the rigid case. 
Figure~\ref{fig:motor_position_grid}(e)--(g) shows that the participating modes are dominated by the rigid body modes $n=0,1$ and the first two bending modes $n=2,3$, with the higher modes $n\geq4$ contributing only weakly.  However the relative amplitude and phase of the excited modes changes with motor position, and is directly responsible for the subtle features observed in figure~\ref{fig:motor_position_grid}(a).

\begin{figure}
  \centering
  \begin{tikzpicture}
    \node[anchor=south west,inner sep=0] (gridimage) at (0,0) {\includegraphics[width=\textwidth]{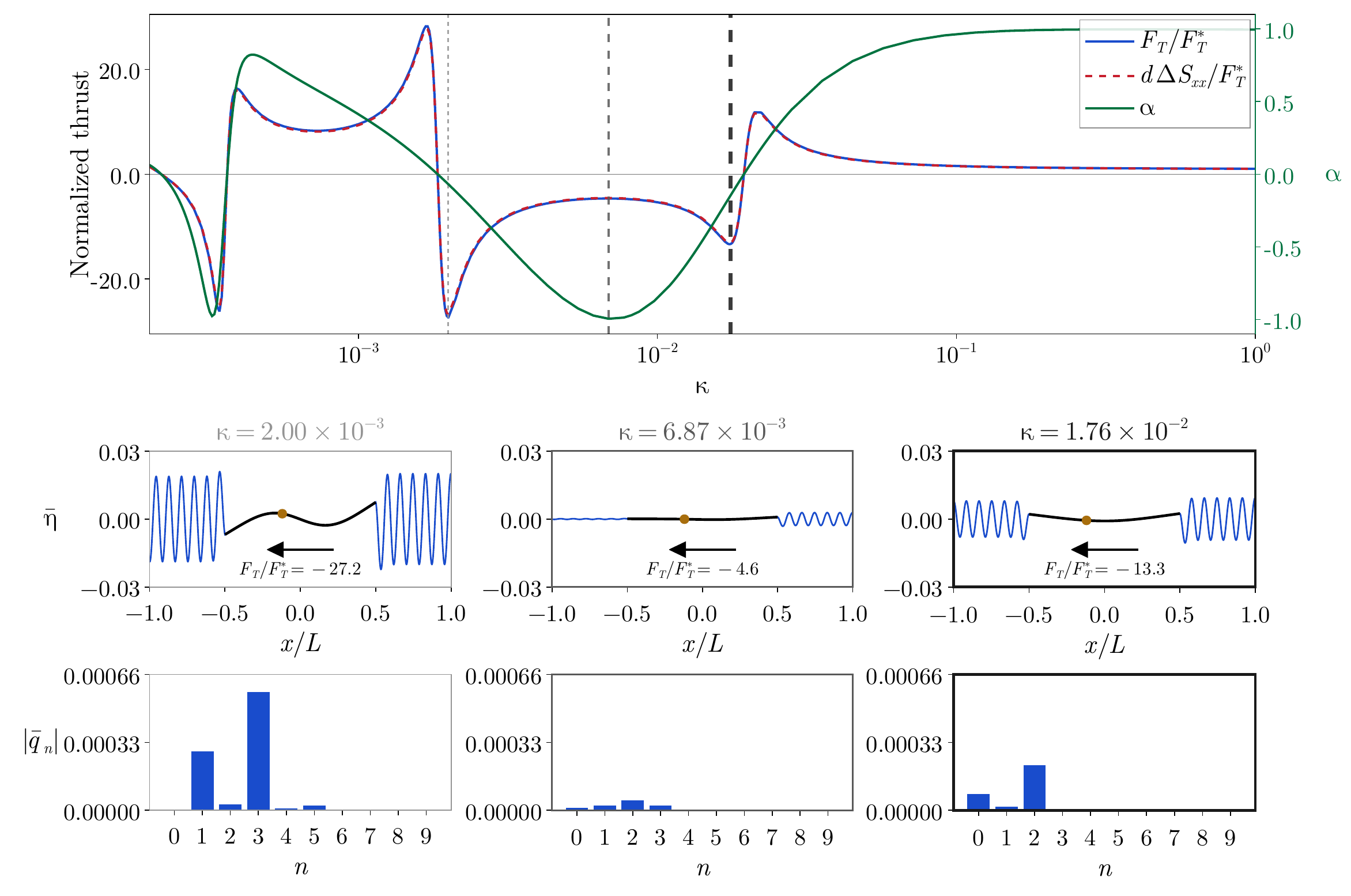}};
    \gridpaneltags
  \end{tikzpicture}
  \caption{Flexibility sweep and representative responses. The top row shows the normalized thrust and the asymmetry factor as a function of non-dimensional stiffness \(\kappa\) at the SurferBot forcing position $x_M/L = -0.12$. Curve colors and line styles are as in figure~\ref{fig:motor_position_grid}; vertical dashed lines mark the three stiffness values shown below, each color and thickness coded to match the corresponding panel below. The middle row shows the corresponding free-surface profiles, with the raft contact segment in black and the motor position marked in gold; the arrow and the labeled value below each raft give the sign and magnitude of \(F_T/F_T^\ast\) at that operating point. The bottom row (e)--(g) shows the modal amplitudes \(|\bar{q}_n|\).}
  \label{fig:flexibility_grid}
\end{figure}

The stiffness sweep in figure~\ref{fig:flexibility_grid}(a) shows the corresponding flexibility effect at the SurferBot forcing position, $x_M/L = -0.12$. Decreasing \(\kappa\) from the rigid limit moves the system through alternating positive and negative thrust regimes. Flexibility also changes the thrust scale, as in this sweep the maximum reaches \(F_T/F_T^\ast\approx28\), more than an order of magnitude above the rigid SurferBot exploration in figure \ref{fig:rigid_position_sweep}. The direct thrust and the Longuet--Higgins prediction track one another closely across the sweep, showing the momentum-flux balance also holds across the entire flexibility sweep. In particular, we see specific isolated ranges of $\kappa$ where the thrust is large and rapidly changes sign.  As analyzed further in what follows, these correspond to regions where a particular elastic mode is near resonance and consequently show a high-amplitude response and rapidly varying phase.

The middle and bottom rows in figure~\ref{fig:flexibility_grid} show the corresponding free-surface profiles and modal mixtures at three select stiffnesses, two at locations of large peak thrust and one at an intermediate case with high asymmetry. The modal amplitudes in panels (e)--(g) of figure~\ref{fig:flexibility_grid} show that the two outer stiffness values are each dominated by a single flexible mode, mode 3 at \(\kappa=2.00\times10^{-3}\) and mode 2 at \(\kappa=1.76\times10^{-2}\), and radiate waves almost symmetrically, whereas the intermediate \(\kappa=6.87\times10^{-3}\) mixes these bending modes with the rigid modes and reaches \(\alpha\approx-1\).  The large amplitude waves and dominant mode shape visualized in (b) and (d) arise from the near resonant excitation of an elastic mode.  Despite being more visibly symmetric (with a low asymmetry factor), these near resonant cases correspond to very large thrust.  To interpret this finding, we return to equation (\ref{eq:SA_decomp_app}) where it was shown $F_T\propto|\bar{\eta}(-\bar{\ell})|^2 - |\bar{\eta}(\bar{\ell})|^2 = -4 \,|S|\, |A| \cos\left(\arg(S) - \arg(A)\right)$.  The resonant excitation of a single elastic mode leads to a large amplitude and rapidly varying phase of either $|S|$ or $|A|$ (depending on whether the resonant $n$ is even or odd, respectively).  As all other modes are away from resonance, their amplitude and phase are relatively insensitive to minor variations in $\kappa$. Therefore, the large maximum thrust magnitude can be directly understood by the large $|S|$ or $|A|$ multiplicative prefactor in equation (\ref{eq:SA_decomp_app}), and the rapid variation in thrust magnitude and sign stems from the strong sensitivity of the corresponding $\arg(S)$ or $\arg(A)$ near resonance.  

From a different perspective, recall that wave thrust is not directly correlated with the difference in wave amplitude ($\Delta \bar{\eta} = \bar{\eta}(\bar{\ell})-\bar{\eta}(-\bar{\ell})$) but rather the difference in their {\it squared} amplitude ($\Delta\overline{\eta^{2}}=|\bar{\eta}(-\bar{\ell})|^{2} - |\bar{\eta}(\bar{\ell})|^{2}$).  As such, even if the wave amplitude difference $\Delta \bar{\eta}$ itself remains unchanged, adding a relatively large but symmetric contribution $\bar{\eta}_n$ to the wave height on both sides (via a proximal resonant mode) leads to a substantial increase in the net momentum flux, via the corresponding increase in $\Delta\overline{\eta^{2}}$ by an amount $\sim\bar{\eta}_n\Delta\bar{\eta}$, to leading order.  However, the introduction of the symmetric resonant contribution $\bar{\eta}_n$ radiates a significant amount of wave energy away that does not directly contribute to thrust production, corresponding to a lower value of $\alpha$. Supplementary Movies 3-5 animate three responses shown in panels (b)--(d).

In fact, the three non-dimensional stiffness values considered here are physically accessible and may directly motivate future experiments. Fixing the SurferBot dimensions (\(5\times3\times0.15\,\mathrm{cm}\)) and forcing frequency $f = 80$ Hz, the corresponding $\kappa$ values are set by the elasticity of the raft material. The values of \(\kappa=2.00\times10^{-3}\), \(6.87\times10^{-3}\), and \(1.76\times10^{-2}\) correspond to Young's moduli of about \(19\), \(67\) and \(171\,\mathrm{MPa}\) respectively, spanning a firm rubber, a rigid polyurethane, and a polyethylene sheet.

The selected cases in figures~\ref{fig:motor_position_grid} and \ref{fig:flexibility_grid} connect points in the one-parameter sweeps to their corresponding wavefields and modal amplitudes. Varying \(x_M/L\) at fixed stiffness changes the relative amplitudes and phases of rigid and bending modes excited by the applied force, but not which modes dominate the response.  Varying \(\kappa\) at a fixed actuator position does change which free--free modes dominate the response, with particularly strong sensitivity in the response when operating near resonance of any single elastic mode.

The snapshots also highlight the relationship between wave directionality and thrust magnitude. The cases with \(\alpha\approx\pm1\) have the strongest left/right imbalance in relative terms, however the largest thrust in the stiffness sweep occurs where the total radiated intensity is larger and a smaller fractional imbalance still produces a larger momentum flux difference. This tradeoff is in fact a common occurrence when analyzing or designing propulsion systems broadly, as high thrust and high efficiency typically occur at different parameters. 

\subsection{Two-dimensional parameter sweep: Thrust}

\begin{figure}
  \centering
  \includegraphics[width=\textwidth]{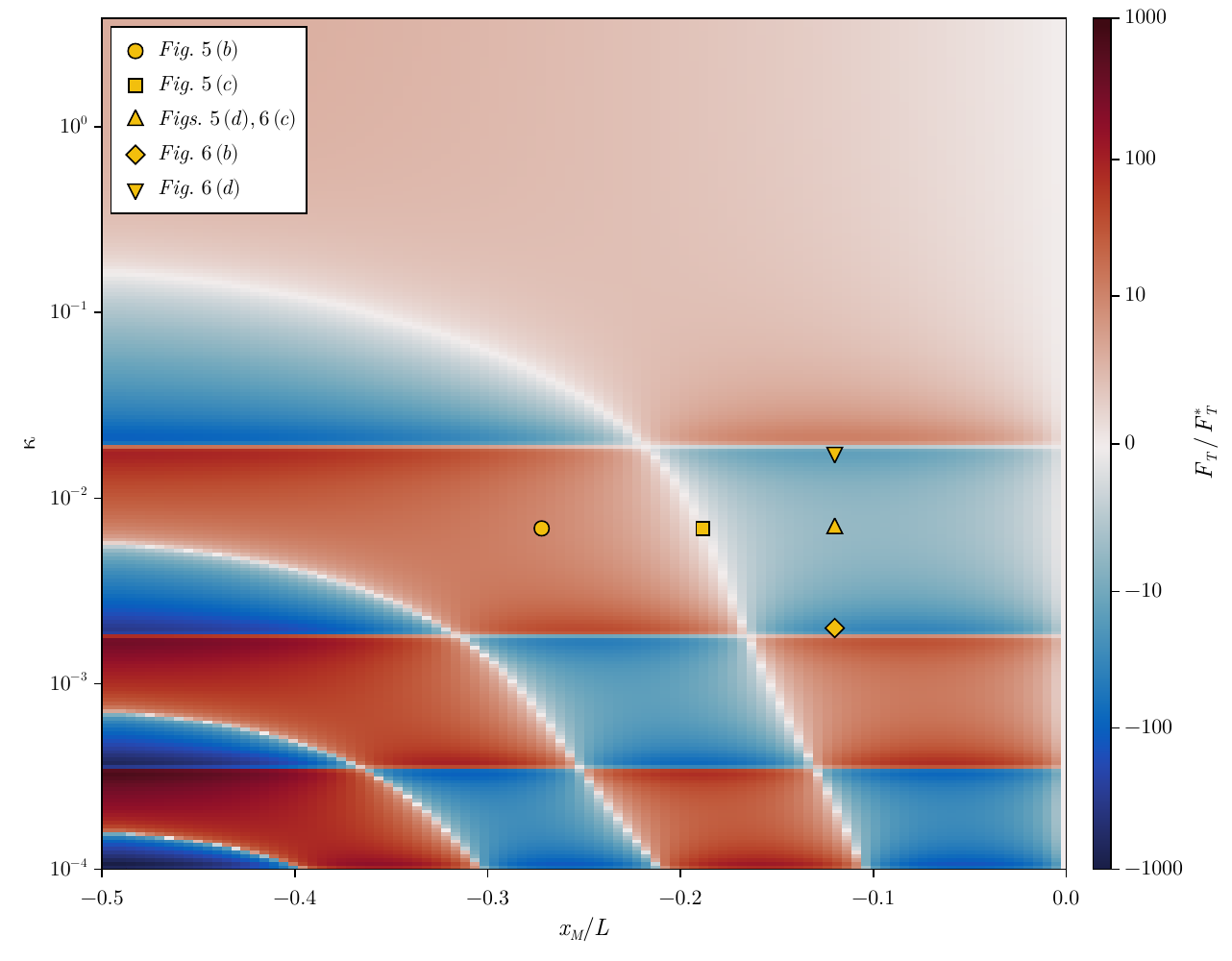}
  \caption{Normalized thrust landscape for the hydrodynamically coupled flexible raft. The color field shows the thrust prediction \(F_T/F_T^\ast\) over the \((x_M/L,\kappa)\) plane. The symbols indicate the specific operating points highlighted in figures~\ref{fig:motor_position_grid} and \ref{fig:flexibility_grid}.}
  \label{fig:thrust_lh_coupled_cbrt}
\end{figure}

Figure~\ref{fig:thrust_lh_coupled_cbrt} generalizes the single parameter slices presented in the prior section to the full \((x_M/L,\kappa)\) plane. The map contains curved patches of positive and negative thrust separated by white boundaries where \(F_T=0\). The superimposed markers place the specific cases highlighted in figures~\ref{fig:motor_position_grid} and \ref{fig:flexibility_grid} within this larger landscape. The one-parameter sweeps in those figures are slices through this full two-parameter response diagram.  The complete thrust landscape in figure~\ref{fig:thrust_lh_coupled_cbrt} shows that stiffness and motor location jointly control the propulsion direction, as visually indicated by the quilt-like appearance of the map. Broad positive and negative regions are separated by sharp transitions, so changing either \(\kappa\) or \(x_M/L\) can reverse the direction of the mean thrust. The large-\(\kappa\) portion at the top of the plot approaches the rigid limit of this non-dimensional sweep. The alternating signed patches and regions of significant thrust enhancement therefore belong to the flexible regime where bending modes carry appreciable displacement. 

We can observe multiple sets of white lines where $F_T = 0$, which correspond to cases where the wavefield is symmetric ($\alpha=0$).  These contours will be interpreted and predicted using the reduced-order model when the corresponding predictions for $\alpha$ are presented in the next section.  The localized patches of high thrust near the horizontal $\alpha=0$ bands correspond to the near resonant cases discussed in the prior section. Explicitly finding parameter combinations that locally maximize thrust could also be identified using the reduced-order model. In particular, one could use equations~\eqref{eq:modal_transfer_rows} and \eqref{eq:modal_transfer_rows2}, and characterize the set of local extrema such that \(\partial F_T/\partial x_M = 0\) and \(\partial F_T/\partial \kappa = 0\) with a second-order condition to delineate maxima/minima/saddles. Such a calculation is left for future work. 

\subsection{Two-dimensional parameter sweep: Asymmetry factor}

\begin{figure}
  \centering
  \includegraphics[width=\textwidth]{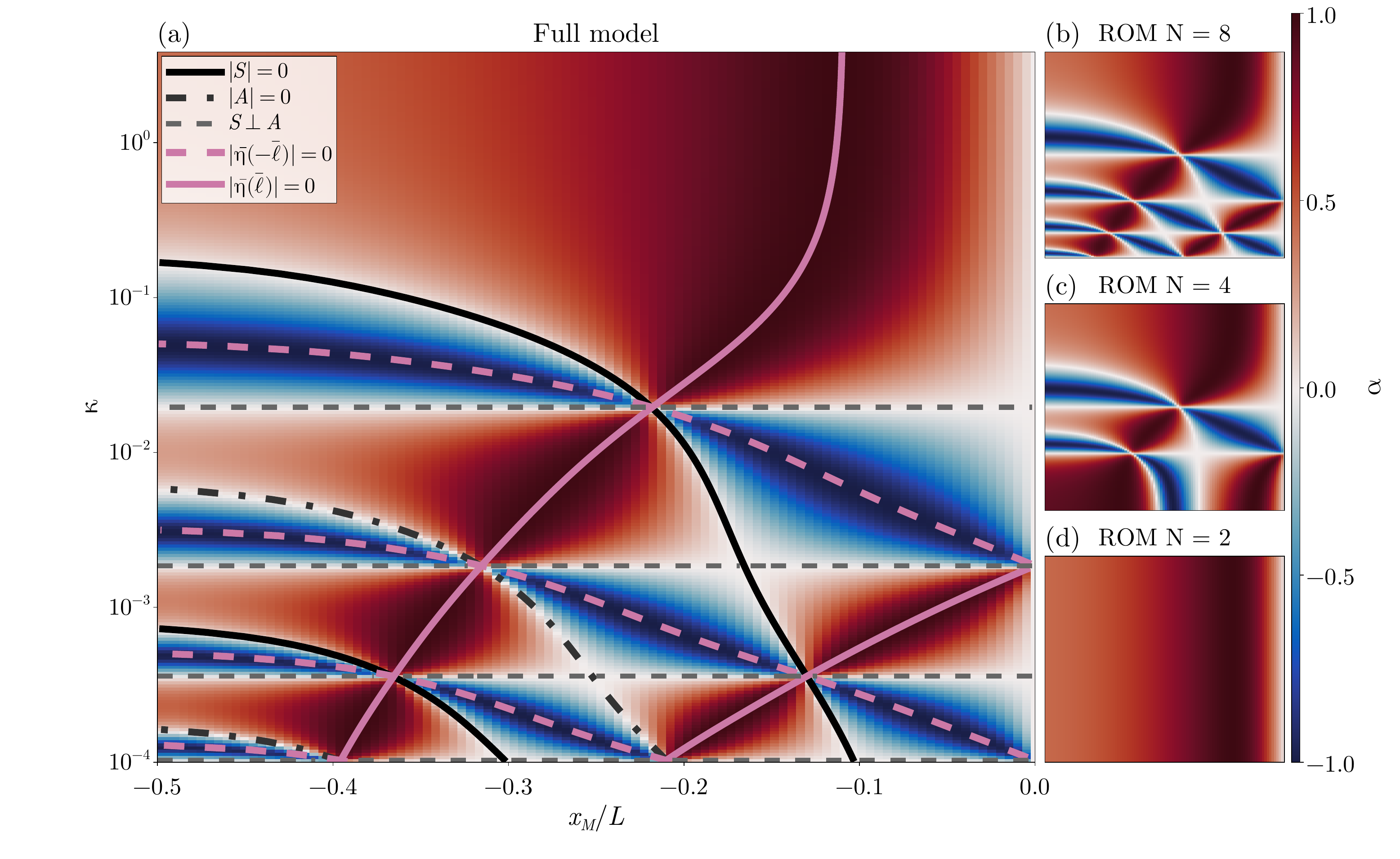}
  \caption{Far-field radiation asymmetry and predictions from modal model. (a) The asymmetry factor \(\alpha\) over the \((x_M/L,\kappa)\) plane from the full model, with overlaid curves obtained from the reduced-order modal radiation map ($N=8$): black curves mark \(\alpha=0\) (solid \(|S|=0\), dash-dot \(|A|=0\), dashed \(S\perp A\)) and magenta curves mark \(\alpha=\pm1\) (\(|\bar\eta(\mp\bar\ell)|=0\)). (b)--(d) The same quantity reconstructed from the modal expansion truncated at \(N=8\), \(N=4\), and \(N=2\) modes, respectively.
  }
  \label{fig:dimensionless_diagnostics}
\end{figure}

Figure~\ref{fig:dimensionless_diagnostics}(a) plots the asymmetry factor \(\alpha\) over the same \((x_M/L,\kappa)\) plane. As in the thrust map, white contours appear that indicate $\alpha=0$ (and thus $F_T=0$) and separate blocks of forward and reverse propulsion. Recalling that both the thrust and asymmetry factor are proportional to $\Real(SA^*)$, each curve ultimately traces one of three conditions: either the symmetric part ($S$) vanishes, the antisymmetric part ($A$) vanishes, or $S \perp A$. These curves can be directly predicted using the reduced-order modal model \eqref{eq:modal_radiation_map}.  To accomplish this, we pre-compute the impedance matrices \(\mathbf{\bar Z}\) and \(\mathbf{\bar C}^\sigma\) once, solve the reduced \(N\times N\) matrix system for \(\bar{\boldsymbol{q}}\), and apply the modal radiation map (Equation~\eqref{eq:modal_radiation_map}) to compute \(S\) and \(A\). Therefore, a single matrix solve provides everything needed to reconstruct a solution, in particular it yields explicit curves for \(|S|=0\), \(|A|=0\), and $S \perp A$, which show excellent agreement when overlaid with the corresponding predictions of the full model on figure~\ref{fig:dimensionless_diagnostics}(a).

The horizontal lines are the loci of $S \perp A$ and are shown as black dashed lines in the figure. In the reduced-order model, the $S\perp A$ condition amounts to finding the roots of a polynomial of degree $N-2$ (Section~\ref{sec:modal-reduction}, Appendix~\ref{app:modal-phase}). This results in predictions for specific values of $\kappa$ at which asymmetry and thrust are both zero, regardless of motor position. The other two families of $\alpha = 0$ curves are the black solid (\(|S|=0\)) and black dashed-dot (\(|A|=0\)) lines.
These three mechanisms produce the alternating band structure apparent in the heatmap and correspond to equal amplitude wave radiation on both sides.  

The same modal model also predicts the curves where $\alpha=\pm1$ (i.e. $|\bar\eta(\pm \bar\ell)| = 0$): the parameter combinations at which the far-field wave amplitude vanishes on one side. These predictions are plotted in magenta and also overlaid on the full model predictions in figure~\ref{fig:dimensionless_diagnostics}(a), once again showing excellent agreement.  As noted prior, locations of maximum asymmetry do not necessarily correspond to maximum thrust, as can be seen by visually comparing the positions of the darkest regions in figures~\ref{fig:dimensionless_diagnostics} and \ref{fig:thrust_lh_coupled_cbrt}.

Figure~\ref{fig:dimensionless_diagnostics}(b)--(d) show the same map reconstructed from the reduced-order modal expansion truncated at $N=8$, $N=4$, $N=2$ modes, respectively. Eight modes reproduce the full model for the range of $\kappa$ explored here without loss of detail. With four modes, the finer structures in the low-$\kappa$ range are not resolved. With only two modes, only the rigid heave and pitch degrees of freedom are retained, so the predicted kinematics are independent of $\kappa$. Nevertheless, as $\kappa \to \infty$, this $N=2$ reduced-order model recovers the rigid solution exactly. More generally, the rigid raft model of \citet{BenhamDevauchelleThomson2024} can be similarly reduced to a two-degrees-of-freedom system using the reduced-order model presented here.

The full heatmap of figure \ref{fig:dimensionless_diagnostics}(a) requires one solve of the discretised coupled system \eqref{eqn:nd_final} at each combination of parameters in the sampled parameter space, about \(3\times10^4\) simulations for the sweep shown here, taking about 84 CPU-hours at roughly a sixth of a CPU-minute per solve. The reduced-order model heatmap of figure~\ref{fig:dimensionless_diagnostics}(b) at $N=8$ instead builds the radiation maps from \(N\) modal basis runs of the full model at about one CPU-minute for the same parameter window, then the full $3\times10^4$ sweep in a few seconds total.  For the parameter regime surveyed here, \(N=8\) showed good convergence.  Extending the exploration to smaller values of $\kappa$ would require inclusion of a larger number of elastic modes in the basis.

\section{Discussion}\label{sec:discussion}

The thrust on a periodically driven floating raft is a result of the fore--aft imbalance of the wave momentum it radiates.
For a rigid raft in two dimensions, the load drives only rigid-body translation and rotation, with actuator placement defining the relative amplitude and phase of those two motions. In this work, we have demonstrated how flexibility adds additional degrees of freedom (flexible modes) for the forcing to excite through a frequency-dependent complex impedance.  Ultimately, effective propulsion requires both
appreciable total radiation and a directional bias. The same two requirements, large availability of wave energy and an asymmetry due to a biased interaction with the structure, govern wave-energy absorption \citep{falnes2015fundamental}, where the target there is absorbed power.

The modal analysis viewpoint applied here to wave-driven propulsion helps reveal why flexibility enables propulsion regimes absent from the rigid problem with otherwise identical forcing. In particular, the non-dimensional flexibility $\kappa$ selects the band of modes that respond most sensitively to forcing, while the actuator position along the raft determines their relative amplitude and phase.  As such, their superposition can lead to new behaviors that were not present in the rigid counterpart, such as a reversal of thrust direction.  In addition, a discrete set of $\kappa$ values emerges in our analysis wherein thrust vanishes for any choice of forcing.  These values occur near structural resonance of an elastic mode, leading to greatly increased wave amplitudes, and large values of thrust in close proximity. Consistent with most other propulsion systems, large thrust does not necessarily result in high efficiency.  As such we additionally characterize the efficiency here through the non-dimensional parameter $\alpha$, which represents the ratio of wave energy that contributes to propulsion to the total radiated wave energy.  A highly efficient use of waves corresponds to $\alpha=\pm1$ wherein waves radiate from a single side only, consistent with the conclusions of the optimal-control analysis of wave-driven propulsion by \citet{ODonovanEtAl2025}.

The alternating block structure of wave asymmetry (and propulsion direction) revealed in our sweep of stiffness and motor position also manifests in the equivalent forced dry (hydrodynamically uncoupled) problem when left and right beam end-point amplitudes are compared (see Appendix~\ref{app:uncoupled-reference}).  Hydrodynamic coupling shifts the locations of these boundaries and determines the far-field momentum flux associated with each structural response, but leaves the underlying branch skeleton visible over the range tested here.  Taking advantage of linearity, we have demonstrated how a modal representation truncated to include $N$ rigid-body and elastic modes can be constructed with only $N$ solves of the full finite-difference model.  The resulting reduced-order model remains fully predictive, with the advantages of being significantly more computationally efficient and physically interpretable.
 
 For a flexible raft driven by waves, material stiffness, actuator placement, and operating frequency are integrally coupled design variables. The same resonance structure organizes single-phase flexible swimming, where thrust peaks at bending rigidities that bring a structural mode into resonance with the forcing \citep{Alben2008,ParazEtAl2016,HooverEtAl2018}. Whether that optimum sits exactly at resonance is debated, with inviscid small-amplitude models placing it at resonance and viscous or strongly nonlinear models placing it off resonance \citep{Wang2022}. Our tethered, inviscid, small-slope calculation falls in the former regime, and the modal reduction shows the thrust peaks occur at the resonances of the underlying structural modes. Furthermore, single-phase flexible propulsion is often characterized by the vortical wake the oscillating body sheds \citep{TriantafyllouEtAl2000,Lauder2015}. For the flexible raft considered here, shed vorticity is absent by construction, and the ``wake'' is characterized solely by the emitted waves.  Interfacial propulsion strategies of small-scale organisms and robots more broadly are characterized by a partition of waves and vortical contributions \citep{Buhler2007}, with the vortical share being dominant for water striders \citep{HuChanBush2003} and the wave effect dominating for the trapped honeybee \citep{RohGharib2019}, for instance.

 The wave-induced drift of ice floes is a close passive counterpart to the problem considered here. \citet{MeylanSquire1996} compute the time-averaged horizontal force on a finite flexible floe from its far-field radiation pattern, and find that the rigid and flexible predictions differ once the floe is compliant enough to bend. Across floes of similar diameter, they also report high sensitivity in the net drift force occurring near resonances of the floe's structural response.

 We conclude with a discussion of the limitations of our work, which also motivate future directions.  As our focus was on elucidating the role of flexibility on wave-driven propulsion, our parameter sweeps varied only $\kappa$ and $x_M/L$, with all other parameters held at the SurferBot reference values.  Nevertheless our underlying mathematical model (and publicly available numerical code) are written in general form so as to enable future explorations of this rich system, and could extend the parameter survey to include the mass ratio $\Gamma$, Froude number $Fr$, Weber number $We$, and width ratio $\Lambda$. In particular, decreasing the length scale $L$ of the raft (and non-dimensional parameters accordingly) will amplify the relative importance of surface tension at the contact line, representing a dynamic form of elastocapillarity typically studied in static or quasi-static scenarios \citep{BicoReyssatRoman2018,RivettiAntkowiak2013,TaroniVella2012}.  Furthermore, the external forcing profile is held at a fixed shape and width in the present study. Alternative spatial distributions, multiple actuator locations, or broadband frequency content may reveal new propulsion regimes. The formulation also allows spatially graded $EI(x)$ and $\rho_R(x)$, but the sweeps presented herein assume uniform beam properties, and as such graded stiffness or density profiles represent an unexplored design consideration in the context of wave-driven propulsion.  The present model is two-dimensional, so effects associated with finite span such as spanwise bending, out-of-plane curvatures, and lateral wave radiation are neglected.  Formal parameter optimization (combining stiffness/mass distribution, actuator profile and location, and operating frequency) and experimental validation of the flexible predictions remain open but are directly enabled by our work. A three-dimensional treatment, in which the raft radiates an omnidirectional wave pattern (such as in the model of \citet{odonovan2026}), is a natural extension of the bidirectional radiation picture developed here and would more closely connect to application. Nevertheless, we anticipate the core qualitative phenomena revealed here to persist, and could be evaluated in experiment with a compliant cousin of the SurferBot.

\begin{acknowledgments}
\noindent\textbf{Supplementary material.} Six movies accompany this paper. Movie 1 shows a raft with non-constant material properties. Movie 2 shows the rigid-limit wavefield of figure~\ref{fig:benham_aguero_validation}(b). Movies 3--5 show the free-surface response at the three stiffnesses of figure~\ref{fig:flexibility_grid}, $\kappa=2.00\times10^{-3}$, $6.87\times10^{-3}$ and $1.76\times10^{-2}$. Movie 6 sweeps the width parameter $\Lambda$ and shows the modal response of figure~\ref{fig:bare_q_uncoupled} evolving as hydrodynamic coupling is introduced. All movies except Movie 2, which is compared against a physical experiment, use non-dimensional coordinates.

\noindent\textbf{Acknowledgements.} The authors thank Jack-William Barotta for fruitful discussions and Jennifer Shim for assistance with the numerical simulations in Python.

\noindent\textbf{Funding.} The authors gratefully acknowledge financial support from the National Science Foundation (CBET-2338320) and the
Office of Naval Research (N00014-21-1-2816 and N00014-21-1-2670).

\noindent\textbf{Competing interests.} The authors declare no competing interests.

\noindent\textbf{Data availability statement.} The numerical code used for the simulations and figure generation is available at \url{https://github.com/harrislab-brown/flexible_surferbot}. 

\noindent\textbf{Use of AI tools.} Claude (Anthropic, Opus 5)
was used during 2026 to copy-edit the manuscript and to verify the analytical results by computer algebra.
No figures were AI-generated or modified, and the authors  have verified and are responsible for all content.

\noindent\textbf{Declaration of interests.} The authors report no conflict of interests.

\end{acknowledgments}

\appendix
\makeatletter
\renewcommand{\theHsection}{app.\Alph{section}}
\renewcommand{\theHequation}{\theHsection.\arabic{equation}}
\makeatother

\section{Mean thrust formula}\label{app:thrust}

This appendix describes how the thrust formula below is calculated from the solution of \eqref{eqn:nd_final}. The mean horizontal thrust on the raft is defined as
\begin{equation}\label{eq:FTbar-def}
F_T
 \;=\;
 \avg{ \int_{-L/2}^{L/2} Q(x,t)\,(\vect{n}(x, t)\!\cdot\!\ex)\,\dd x }
 \;+\; \sigma\,d\,\Bigl[\avg{\cos\theta}\Bigr]_{x=-L/2}^{x=L/2}.
\end{equation}
Here \(Q=p \, d - f\) is the net vertical load exerted on the raft per unit streamwise length, with positive \(f\) acting downward and positive \(\eta\) upward.

For small slopes, the capillary contribution gives \(\avg{\cos\theta} = 1 - \tfrac{1}{4}\,|\hat{\eta}_x|^2\). Substituting into \eqref{eq:FTbar-def} yields
\begin{equation}\label{eq:FTbar2}
F_T
 \;=\;
 \int_{-L/2}^{L/2} \avg{\,Q(x,t)\,(\vect{n}\!\cdot\!\ex)\,}\,\dd x
 \;+\; \sigma\,d\,\Bigl[1 - \tfrac{1}{4}\,|\hat{\eta}_x|^2\Bigr]_{x=-L/2}^{x=L/2}.
\end{equation}
We then approximate the projection in terms of the free-surface slope as
\[
(\vect{n}\!\cdot\!\ex) \approx -\eta_x
  = -\Real\!\left\{ \hat{\eta}_x\,\mathrm{e}^{i\omega t} \right\}
\]
which leads to
\begin{align}
F_T
&= -\int_{-L/2}^{L/2}
    \avg{\,\Real\!\{(\hat{p} \,d - \hat{f})\,\mathrm{e}^{i\omega t}\}\;
          \Real\!\{\hat{\eta}_x\,\mathrm{e}^{i\omega t}\}\,}\,\dd x
   \;+\; \sigma\,d\,\Bigl[1 - \tfrac{1}{4}\,|\hat{\eta}_x|^2\Bigr]_{x=-L/2}^{x=L/2}.
\end{align}
Lastly, using the standard identity
\begin{equation} \label{eqn:complex_identity}
\avg{\,\Real\{\hat{a}\mathrm{e}^{i\omega t}\}\,
        \Real\{\hat{b}\mathrm{e}^{i\omega t}\}\,}
= \frac{1}{2} \left( \Real \{ \hat{a}^* \, \hat{b}\} \right)
= \tfrac{1}{2}\!\left( \Real\{\hat{a}\}\Real\{\hat{b}\}
                      + \Imag\{\hat{a}\}\Imag\{\hat{b}\} \right),
\end{equation}
we obtain
\begin{align}
F_T
= -&\frac{1}{2}\int_{-L/2}^{L/2}
     \Big(
       \Real\{\hat{p}\,d - \hat{f}\}\,\Real\{\hat{\eta}_x\}
       + \Imag\{\hat{p}\,d - \hat{f}\}\,\Imag\{\hat{\eta}_x\}
     \Big)\,\dd x \\
   &+ \sigma\,d\,\Bigl[1 - \tfrac{1}{4}\,|\hat{\eta}_x|^2\Bigr]_{x=-L/2}^{x=L/2}. \nonumber
\end{align}
This is the expression evaluated as the direct thrust in the numerical results. The far-field radiation stress \eqref{eq:Sxx_farfield_results} provides the independent momentum check for the inviscid limit. The integral is evaluated numerically using the trapezoidal rule.

\section{Modal projection and discrete stiffnesses}\label{app:added-mass}

This appendix derives the modal projection \eqref{eq:q_modal_W_app} and details the construction of the hydrodynamic impedance matrix $\bar{\mathbf Z}$ \eqref{eq:unit_Wn_prescription_app}, the capillary endpoint map $\bar{\mathbf C}^\sigma$ \eqref{eq:Ksigma_linear_map_app}, and the modal radiation map \eqref{eq:modal_radiation_map}. It also further expands on the even/odd symmetric structure of the problem introduced in the main text \eqref{eq:modal_parity_blocks}, which allows for a more direct interpretation of how the model outputs relate to thrust production.

\subsection{Free--free modal projection}\label{app:freefree-modes}
We work with $L^2([-1/2,1/2])$-normalized free--free Euler--Bernoulli modes satisfying
\begin{equation}
W_n''''=\beta_n^4W_n,\quad
W_n''(\pm 1/2)=W_n'''(\pm 1/2)=0,
\quad
\int_{-1/2}^{1/2}W_mW_n\,\dd x=\delta_{mn}.
\label{eq:free_free_modes_app}
\end{equation}

\begin{figure}
  \centering
  \includegraphics[width=0.9\textwidth]{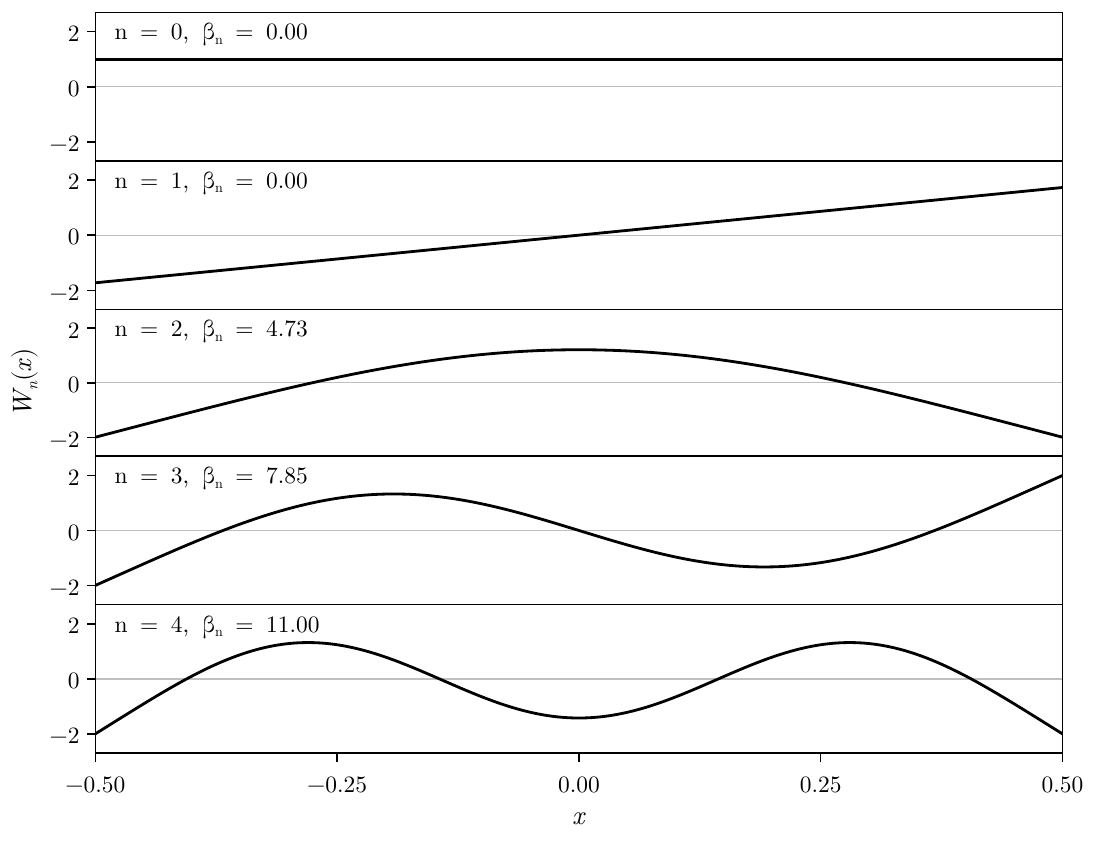}
  \caption{$L^2$-normalized free--free Euler--Bernoulli modes $W_0$--$W_4$.
           Modes $n=0$ and $n=1$ are the rigid-body translation and rotation;
           modes $n\geq2$ are the elastic bending shapes.}
  \label{fig:freefree_modes}
\end{figure}

\noindent The rigid-body modes are $W_0(x)=1$ and $W_1(x)=\sqrt{12}\,x$. For elastic modes $n\geq2$, $W_n=\mathcal N_n\widetilde W_n$, where $\mathcal N_n$ is a normalization factor chosen so that $\int_{-1/2}^{1/2}W_n^2\,\dd x=1$. With the origin at the beam centre,
\begin{align}
\widetilde W_n(x)
&=\frac{\cosh(\beta_nx)}{\cosh(\beta_n/2)}
  +\frac{\cos(\beta_nx)}{\cos(\beta_n/2)},
&&n \geq 2\ \mathrm{even},\\
\widetilde W_n(x)
&=\frac{\sinh(\beta_nx)}{\sinh(\beta_n/2)}
  +\frac{\sin(\beta_nx)}{\sin(\beta_n/2)},
&&n \geq 2\ \mathrm{odd}.
\end{align}
The corresponding eigenvalues satisfy the transcendental equation $\tan(\beta_n/2)=-\tanh(\beta_n/2)$ for even elastic modes and $\tan(\beta_n/2)=\tanh(\beta_n/2)$ for odd elastic modes \citep{rao1995mechanical}. Figure~\ref{fig:freefree_modes} shows $W_0$ through $W_4$ on $x\in[-1/2,1/2]$ along with their corresponding values of $\beta_n$, where we note that $
W_n(-x) = (-1)^n W_n(x)$.

The orthonormal modes $W_n$ form a complete basis for $L^2([-1/2,1/2])$. We start with the non-dimensional raft force balance \eqref{eq:nd_beam_modal_start},
\begin{equation}
\kappa\bar{\eta}_{xxxx}
+\left(\frac{\Lambda\Gamma}{Fr^2}-1\right)\bar{\eta}
=\Lambda\Gamma\,\bar{p}_{\mathrm{dyn}}-\bar{f},
\qquad |x|\le 1/2.
\label{eq:nd_beam_modal_start_app}
\end{equation}
We can project any function of interest onto free-free modal basis, for example
\begin{equation}
\bar\eta(x) = \sum_{n\geq 0} \bar{q}_n W_n(x),  \quad \bar q_n=\int_{-1/2}^{1/2}\bar\eta(x)W_n(x)\,\dd x.
\label{eq:q_modal_W_app}
\end{equation}
We then project all terms of equation \eqref{eq:nd_beam_modal_start_app} onto these modes. The inertia and hydrostatic terms project immediately:
\begin{equation}
\int_{-1/2}^{1/2}W_m
\left(\frac{\Lambda\Gamma}{Fr^2}-1\right)\bar\eta\,\dd x
=\left(\frac{\Lambda\Gamma}{Fr^2}-1\right)\bar q_m.
\label{eq:mass_hydro_proj_app}
\end{equation}
For the bending term, four integrations by parts give
\begin{equation}
\begin{split}
\int_{-1/2}^{1/2}W_m\kappa\bar\eta_{xxxx}\,\dd x
&=\kappa\beta_m^4\bar q_m\\
&\quad+\kappa\left[W_m\bar\eta_{xxx}-W_m'\bar\eta_{xx}
 +W_m''\bar\eta_x-W_m'''\bar\eta\right]_{-1/2}^{1/2}.
\end{split}
\label{eq:ibp_app}
\end{equation}
The term involving $\bar\eta_{xx}$ vanishes since the raft moment is zero at $x=\pm1/2$ due to \eqref{eq:moment_zero}, while $W_m''$ and $W_m'''$ vanish because the basis functions satisfy the free--free boundary conditions \eqref{eq:free_free_modes_app}. The remaining endpoint term is instead fixed by the capillary traction balance \eqref{eq:shear_bc_dim}, since \eqref{eq:q_modal_W_app} converges in $L^2$ and cannot be differentiated termwise, giving
\begin{equation}
\kappa W_m\bar\eta_{xxx}\Bigr|_{-1/2}^{1/2}
=\frac{\Lambda}{We}\left[W_m(1/2)\bar\eta_x(1/2^+)
 +W_m(-1/2)\bar\eta_x(-1/2^-)\right]
=\frac{\Lambda}{We}\bar K_m^\sigma,
\label{eq:capillary_boundary_stiffness_app}
\end{equation}
where we recall $\bar{\eta}_x(1/2^+):= \lim_{y\downarrow1/2}\bar{\eta}_x(y)$, and similarly for $\bar{\eta}_x(-1/2^-)$. Substitution gives the modal balance \eqref{eq:modal_balance_adim_app}. If we define $s_n^+=\bar\eta_x(1/2^+)|_{\bar\eta=W_n}$ and $s_n^-=\bar\eta_x(-1/2^-)|_{\bar\eta=W_n}$, they can be recorded from the solve forced by mode $W_n$ as
\begin{equation}
\bar K_m^\sigma=\sum_{n=0}^{N-1}\bar C_{mn}^\sigma\bar q_n,
\qquad
\bar C_{mn}^\sigma=W_m(1/2)s_n^+ +W_m(-1/2)s_n^-.
\label{eq:Ksigma_linear_map_app}
\end{equation}
Thus the capillary endpoint map can be assembled alongside the hydrodynamic impedance matrix and appears in the modal matrix as the $(\Lambda/We)\bar{\mathbf C}^\sigma$ contribution.

Similarly, the entries of the hydrodynamic impedance matrix $\bar{\mathbf Z}$ are obtained by evaluating the fluid map on the basis functions themselves. For the $n$th column, we prescribe the wetted displacement by setting
\begin{equation}
\bar\eta(x)=W_n(x),
\qquad -1/2\leq x\leq1/2,
\label{eq:unit_Wn_prescription_app}
\end{equation}
solve the corresponding linear fluid problem \eqref{eqn:nd_final} numerically, reconstruct $\bar p_{\mathrm{dyn}}$, and project it through \eqref{eq:modal_load_definitions}. Since the prescribed modal vector is the $n$th unit vector, the projected loads are the entries of the $n$th column of $\bar{\mathbf Z}$. In the discretized domain implementation, the vectors $W_n(x_i)$ are evaluated on the raft grid and are orthogonal up to some rounding error which grows with $n$, thus the Gram matrix \citep{golub2013matrix} of the sampled basis is used to recover orthonormality of the modal amplitudes in the finite-dimensional formulation.

\subsection{Parity structure of the modal system}\label{app:modal-parity}

Within the present study, the centered raft and fluid domain have left--right symmetry. This subsection shows why the underlying matrix problem inherits the same split and its implications. A displacement $\bar\eta = W_n$ produces a dynamic pressure \(\bar p_{\mathrm{dyn}}^{(n)} = \bar p_{\mathrm{dyn}}\big|_{\bar \eta = W_n}\) with the same parity, $\bar p_{\mathrm{dyn}}^{(n)}(-x)=(-1)^n\bar p_{\mathrm{dyn}}^{(n)}(x)$. The entry in row $m$ and column $n$ of the pressure map is
\begin{equation}
\bar Z_{mn}=\int_{-1/2}^{1/2}W_m(x)\bar p_{\mathrm{dyn}}^{(n)}(x)\,\dd x
=(-1)^{m+n}\bar Z_{mn}.
\label{eq:modal_impedance_parity}
\end{equation}
It follows that $\bar Z_{mn}=0$ when $m$ and $n$ have opposite parity. Reflection exchanges the two edge slopes in \eqref{eq:Ksigma_linear_map_app}, with the sign set by the parity of $W_n$, so $\bar{\mathbf C}^\sigma$ has the same property, as well as the diagonal matrix $\bar{\mathbf D}$. These observations give the block system \eqref{eq:modal_parity_blocks}. Parity determines which matrix entries vanish. 

During the solve with $\bar\eta=W_n$ prescribed on the raft, we also store the boundary amplitudes $\bar a_n^+=\bar\eta(\bar\ell)|_{W_n}$ and $\bar a_n^-=\bar\eta(-\bar\ell)|_{W_n}$. For any modal response $\bar{\boldsymbol q}$, the far-field amplitudes are
\begin{equation}
\bar\eta(\pm\bar\ell)=\sum_{n=0}^{N-1}\bar a_n^\pm\bar q_n.
\end{equation}
For the symmetric raft and computational domain, we have $\bar a_n^-=(-1)^n\bar a_n^+$. Writing $\bar a_n\equiv\bar a_n^+$, this becomes
\begin{equation}
\bar\eta(\bar\ell)=\sum_{n=0}^{N-1}\bar a_n\bar q_n,
\qquad
\bar\eta(-\bar\ell)=\sum_{n=0}^{N-1}(-1)^n \bar a_n\bar q_n.
\label{eq:boundary_modal_app}
\end{equation}
Substituting into the definitions of $S$ and $A$ gives
\begin{equation}
S=\sum_{n\ \mathrm{even}}\bar a_n\bar q_n = \boldsymbol{\bar a}_e^{\mathsf T}\bar{\boldsymbol q}_e,
\qquad
A=\sum_{n\ \mathrm{odd}}\bar a_n\bar q_n = \boldsymbol{\bar a}_o^{\mathsf T}\bar{\boldsymbol q}_o.
\label{eq:SA_modal_app}
\end{equation}
A response built from even modes alone therefore radiates the same amplitude to both sides, $\bar\eta(\bar\ell)=\bar\eta(-\bar\ell)=S$, and one built from odd modes alone radiates equal and opposite amplitudes, $\bar\eta(\bar\ell)=-\bar\eta(-\bar\ell)=A$.

\subsection{Zero-thrust condition}\label{app:modal-phase}

This subsection proves that stiffness and motor position decouple and act independently on thrust, a claim used throughout Section \ref{sec:results} and Section \ref{sec:discussion}. We will also see that an energy conservation argument leads to a common-phase structure that explains the horizontal lines in figures \ref{fig:thrust_lh_coupled_cbrt} and \ref{fig:dimensionless_diagnostics}(a). We define the real and imaginary parts  in \eqref{eq:modal_radiation_map} by
\begin{equation}
\bar{\mathbf M}_p=\bar{\mathbf H}_p+i\bar{\mathbf Y}_p,
\label{eq:modal_reactive_radiative_parts}
\end{equation}
where $p\in \{e, o\}$, and $\bar{\mathbf H}_p$ and $\bar{\mathbf Y}_p$ are symmetric real matrices \citep{korobkinEigenmodesAddedmassMatrices2023}. The matrix $\bar{\mathbf H}_p$ contains the elastic, inertial, hydrostatic, capillary, and reactive (added mass) fluid forces, all of which exchange energy with the raft over a cycle but have zero mean power.

The matrix $\bar{\mathbf Y}_p$ accounts for the mean energy carried away by the outgoing waves, and we will see that it has rank one. To see this directly, prescribe any complex modal displacement $\bar{\boldsymbol q}_p$. The mean power the actuator delivers to that motion can be calculated as
\begin{align}
\left\langle P_{\mathrm{in}}\right\rangle
&=-\left\langle\int_{-1/2}^{1/2} f\,\eta_{t}\dd x\right\rangle
  \label{eq:modal_power_physical}\\
&=-\left\langle\int_{-1/2}^{1/2}
   \Bigl(\sum_{m\in p}\Real\{\bar f_{m}\mathrm{e}^{it}\}W_{m}\Bigr)
   \Bigl(\sum_{n\in p}\Real\{i\bar q_{n}\mathrm{e}^{it}\}W_{n}\Bigr)\dd x\right\rangle
  \nonumber\\
&=-\sum_{m,n\in p}
   \left\langle\Real\{\bar f_{m}\mathrm{e}^{it}\}\,\Real\{i\bar q_{n}\mathrm{e}^{it}\}\right\rangle
   \int_{-1/2}^{1/2}W_{m}W_{n}\dd x
  \nonumber\\
&=-\sum_{n\in p}
   \left\langle\Real\{\bar f_{n}\mathrm{e}^{it}\}\,\Real\{i\bar q_{n}\mathrm{e}^{it}\}\right\rangle
  \nonumber\\
&=-\frac{1}{2}\sum_{n\in p}\Real\{\bar f_{n}^{*}\,i\bar q_{n}\}
 =-\frac{1}{2}\Real\!\left\{(\bar{\boldsymbol f}_{p}^{*})^{\mathsf T}i\bar{\boldsymbol q}_{p}\right\}
  \nonumber\\
&=\frac{1}{2}\Real\!\left\{
   \left[(\bar{\mathbf M}_{p}\bar{\boldsymbol q}_{p})^{*}\right]^{\mathsf T}
   i\bar{\boldsymbol q}_{p}\right\}
  \nonumber\\
&=\frac{1}{2}\Real\!\left\{
   i\,(\bar{\boldsymbol q}_{p}^{*})^{\mathsf T}\bar{\mathbf H}_{p}^{\mathsf T}\bar{\boldsymbol q}_{p}
   +(\bar{\boldsymbol q}_{p}^{*})^{\mathsf T}\bar{\mathbf Y}_{p}^{\mathsf T}\bar{\boldsymbol q}_{p}
   \right\}
  \nonumber\\
&=\frac{1}{2}(\bar{\boldsymbol q}_{p}^{*})^{\mathsf T}\bar{\mathbf Y}_{p}\bar{\boldsymbol q}_{p},
  \label{eq:modal_input_power}
\end{align}
where we used the facts that the set $W_n$ forms an orthonormal basis, the relation \eqref{eqn:complex_identity}, and that the term containing $\bar{\mathbf H}_p$ is purely imaginary inside the real part and therefore contributes no net cycle-averaged power.  The latter result corresponds to the fact that the physical effects captured in $\bar{\mathbf H}_p$ act $\pi/2$ out of phase with the velocity. Due to conservation of wave energy, \eqref{eq:modal_input_power} must equal the outgoing wave energy:
\begin{equation}
\left\langle P_{\mathrm{rad}}\right\rangle
:=\bar J_p |\bar\eta(\pm\bar\ell)|^2 = \bar J_p |\boldsymbol{\bar a}_p^{\mathsf T}
\bar{\boldsymbol q}_p|^2 = \bar J_p(\bar{\boldsymbol q}_p^*)^{\mathsf T}
\boldsymbol{\bar a}_p^*\boldsymbol{\bar a}_p^{\mathsf T}
\bar{\boldsymbol q}_p.
\label{eq:modal_radiated_power}
\end{equation}
Here, let $\bar J_p$ be the nondimensional wave-energy flux associated with unit outgoing amplitude. With no viscous dissipation ($\nu = 0$), input and radiated power are equal for every prescribed $\bar{\boldsymbol q}_p$:
\begin{equation}
\bar{\mathbf Y}_p=2\bar J_p
\boldsymbol{\bar a}_p^*\boldsymbol{\bar a}_p^{\mathsf T}.
\label{eq:modal_rank_one_radiation}
\end{equation}
Since $\bar{\mathbf Y}_p$ is real, $\bar a_{p,m}^*\bar a_{p,n}$ is real for every pair of modes. All nonzero entries of $\boldsymbol{\bar a}_p$ therefore share one complex phase, with signs contained in a real vector:
\begin{equation}
\boldsymbol{\bar a}_p=\mathrm e^{i\theta_p}\boldsymbol{\bar c}_p,
\qquad \boldsymbol{\bar c}_p\in\mathbb R^{n_p}.
\label{eq:modal_radiation_direction}
\end{equation}
Writing $\gamma_p=2\bar J_p$, the modal block becomes
\begin{equation}
\bar{\mathbf M}_p
=\bar{\mathbf H}_p+i\gamma_p\boldsymbol{\bar c}_p\boldsymbol{\bar c}_p^{\mathsf T}.
\label{eq:modal_rank_one_block}
\end{equation}
The Sherman--Morrison formula \citep{golub2013matrix} gives the inverse of a rank-one matrix sum in \eqref{eq:modal_transfer_rows} as
\begin{equation}
\boldsymbol{\bar r}_p
=-\boldsymbol{\bar a}_p^{\mathsf T}\bar{\mathbf M}_p^{-1}
=-\mathrm e^{i\theta_p}
\frac{\boldsymbol{\bar c}_p^{\mathsf T}\bar{\mathbf H}_p^{-1}}
{1+i\gamma_p\boldsymbol{\bar c}_p^{\mathsf T}\bar{\mathbf H}_p^{-1}\boldsymbol{\bar c}_p}.
\label{eq:modal_transfer_phase_formula}
\end{equation}
This expression assumes that $\bar{\mathbf H}_p$ is invertible. The adjugate form derived below remains valid if $\bar{\mathbf H}_p$ is singular but $\bar{\mathbf M}_p$ is invertible. The numerator in \eqref{eq:modal_transfer_phase_formula} is a real row, and all of its entries have the same complex denominator. Hence
\begin{equation}
\boldsymbol{\bar r}_p(\kappa)
=\mathrm e^{i\delta_p(\kappa)}\boldsymbol{\bar b}_p(\kappa)^{\mathsf T},
\qquad \boldsymbol{\bar b}_p(\kappa)\in\mathbb R^{n_p}.
\label{eq:modal_common_phase}
\end{equation}
Both $\boldsymbol{\bar b}_p$ and $\delta_p$ generally change with stiffness. Substitution into \eqref{eq:modal_interference_matrix} gives
\begin{equation}
\mathbf G(\kappa)
=\cos\!\left(\delta_e(\kappa)-\delta_o(\kappa)\right)
\boldsymbol{\bar b}_e(\kappa)\boldsymbol{\bar b}_o(\kappa)^{\mathsf T}.
\label{eq:modal_G_common_phase}
\end{equation}
At a stiffness where both transfer rows are nonzero, $\mathbf G(\kappa)=\mathbf0$ when $\delta_e-\delta_o=\pi/2$ modulo $\pi$. Equation~\eqref{eq:modal_load_family_condition} then gives $\Real(SA^*)=0$ for every real forcing profile. If $S$ and $A$ are both nonzero,
\begin{equation}
\Real(SA^*)=|S||A|\cos\!\left(\arg S-\arg A\right)=0
\end{equation}
is precisely the condition $S\perp A$. If either amplitude vanishes, the response instead belongs to the $S=0$ or $A=0$ family and their relative phase is undefined.

\subsubsection{Out-of-phase condition}\label{app:out-of-phase}
It remains to calculate the stiffnesses at which this phase difference occurs. The previous subsection reduced $\mathbf G(\kappa)=\mathbf 0$, away from the degenerate $S=0$ or $A=0$ branches, to the single condition $\delta_e(\kappa)-\delta_o(\kappa)=\pi/2$ modulo $\pi$, so we now track how the phases $\delta_e(\kappa)$ and $\delta_o(\kappa)$ depend on $\kappa$. At fixed geometry and fluid parameters, $\kappa$ enters only through the bending term. Define
\begin{equation}
\bar{\mathbf B}_p=\operatorname{diag}(\beta_n^4)_{n\in p},
\qquad
\bar{\mathbf H}_p(\kappa)=\bar{\mathbf H}_{p,0}+\kappa\bar{\mathbf B}_p,
\label{eq:modal_affine_reactive_block}
\end{equation}
where $\bar{\mathbf H}_{p,0}$ contains all real terms that remain at $\kappa=0$. We define the determinant of the modal block as $P_p(\kappa)$. Applying the matrix determinant lemma \citep{harville1997matrix} to the rank-one radiation term in \eqref{eq:modal_rank_one_block} gives
\begin{align}
P_p(\kappa)
&=\det\bar{\mathbf M}_p(\kappa)\\
&=\det\bar{\mathbf H}_p(\kappa)
+i\gamma_p\boldsymbol{\bar c}_p^{\mathsf T}
\operatorname{adj}(\bar{\mathbf H}_p(\kappa))\boldsymbol{\bar c}_p.
\label{eq:modal_denominator_polynomials}
\end{align}
Substituting $\bar{\mathbf H}_p^{-1}=\operatorname{adj}(\bar{\mathbf H}_p)/\det(\bar{\mathbf H}_p)$ into \eqref{eq:modal_transfer_phase_formula} and using \eqref{eq:modal_denominator_polynomials} to rewrite its denominator, $\det(\bar{\mathbf H}_p)$ cancels and the response to a real forcing vector becomes
\begin{equation}
\boldsymbol{\bar a}_p^{\mathsf T}\bar{\mathbf M}_p^{-1}\bar{\boldsymbol f}_p
=\mathrm e^{i\theta_p}
\frac{\boldsymbol{\bar c}_p^{\mathsf T}
\operatorname{adj}(\bar{\mathbf H}_p)\bar{\boldsymbol f}_p}
{P_p(\kappa)}.
\label{eq:modal_adjugate_response}
\end{equation}
The numerator of \eqref{eq:modal_adjugate_response}, $\boldsymbol{\bar c}_p^{\mathsf T}\operatorname{adj}(\bar{\mathbf H}_p)\bar{\boldsymbol f}_p$, is real, since $\boldsymbol{\bar c}_p$, $\operatorname{adj}(\bar{\mathbf H}_p)$ and $\bar{\boldsymbol f}_p$ are all real. The phase is thus fixed by $\theta_p$ and $P_p(\kappa)$ alone. The out-of-phase condition $\delta_e(\kappa)-\delta_o(\kappa)=\pi/2$ modulo $\pi$ is therefore equivalent to finding the positive roots of the single real polynomial
\begin{equation}
\mathcal P(\kappa)
=\Real\!\left[
\mathrm e^{i(\theta_e-\theta_o)}P_e(\kappa)^*P_o(\kappa)
\right]=0,
\label{eq:modal_kappa_polynomial}
\end{equation}
which is a polynomial in $\kappa$ because $P_e(\kappa)$ and $P_o(\kappa)$ are, as $\bar{\mathbf H}_p$ is affine in $\kappa$, so both $\det\bar{\mathbf H}_p(\kappa)$ and $\operatorname{adj}(\bar{\mathbf H}_p(\kappa))$ are polynomials in $\kappa$. At each root of \eqref{eq:modal_kappa_polynomial}, $S\perp A$ for every motor position, independent of how the raft is forced.

\begin{figure}
  \centering
  \includegraphics[width=0.95\textwidth]{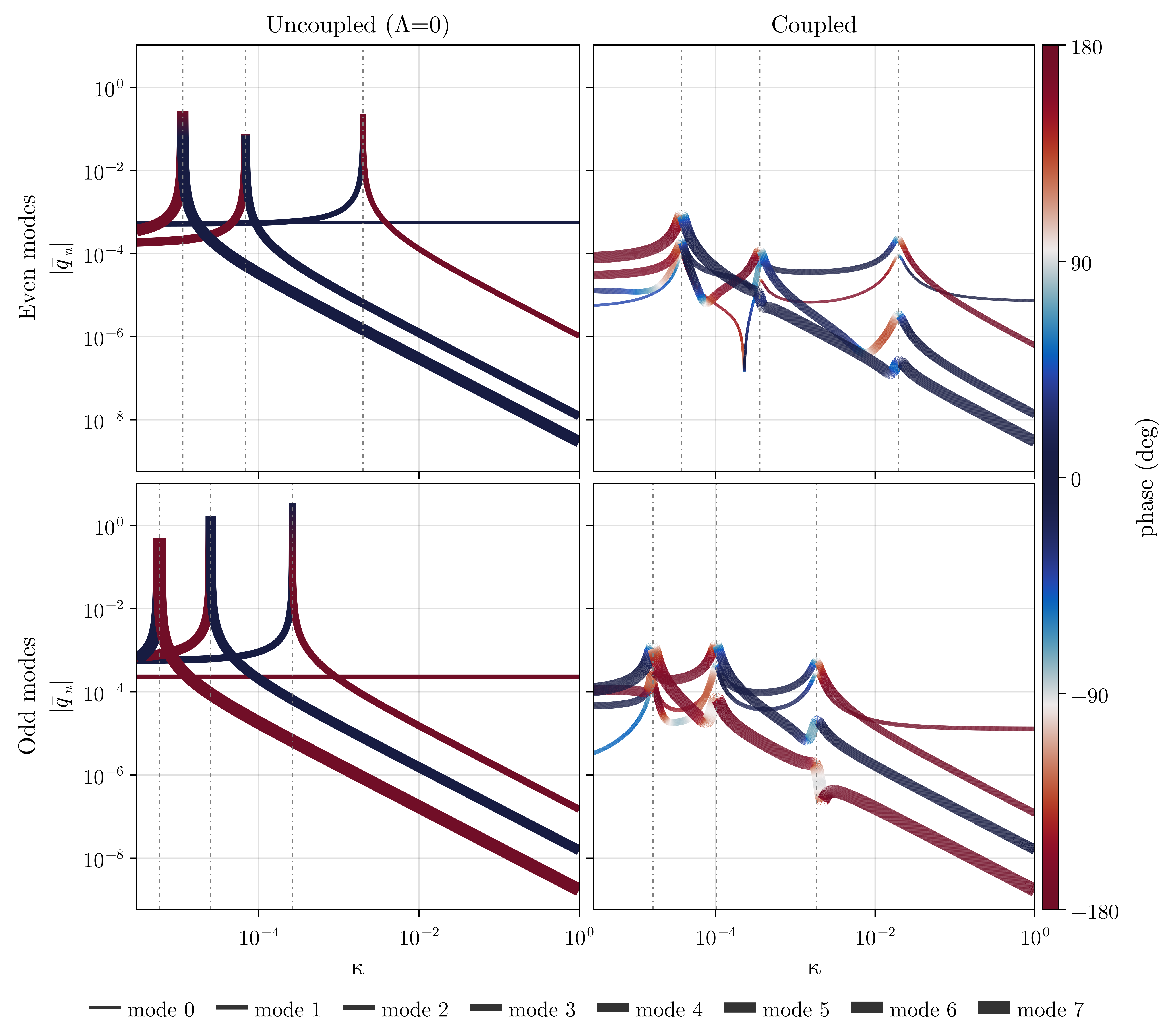}
  \caption{Individual modal response \(|\bar q_n|\) versus \(\kappa\), split by parity (rows) and by hydrodynamic coupling (columns). Color encodes the phase of $q_n$ and line thickness encodes mode number. Vertical dash-dotted lines mark resonant $\kappa$ values, calculated numerically as the zeroes of $\mathbf{G}(\kappa)$.  A video where the plot is shown as $\Lambda$ is gradually increased from 0 is provided as Supplementary Movie 6.}
  \label{fig:bare_q_uncoupled}
\end{figure}

\section{Uncoupled limit $\Lambda = 0$} \label{app:uncoupled-reference}

Equation~\eqref{eq:modal_radiation_map} reduces to the hydrodynamically uncoupled beam response when the width parameter is set to \(\Lambda=d/L=0\). In this limit the pressure loading and capillary endpoint terms (proportional to \(d\)) are removed, so the raft response is governed by a balance of only the imposed forcing with the intrinsic beam inertia and stiffness. This limit recovers the anticipated modal geometry of the free--free beam.  Figure~\ref{fig:bare_q_uncoupled} shows the bare modal amplitudes \(|\bar q_n|\) versus \(\kappa\), split by parity, comparing the uncoupled limit against the fully coupled case. In the uncoupled column, each elastic mode sharply peaks at its own resonance and decays away from it. Since this limit is undamped, the phase sits exactly at $0^{\circ}$ or $\pm 180^{\circ}$ on either side of a resonance, jumping discretely between the two at the pole. The rigid modes ($n=0, 1$) are insensitive to $\kappa$ altogether, since $\beta_0=\beta_1=0$. In the coupled column, the resonant $\kappa$ values shift, and modes of the same parity respond at each of them, including the rigid ones, since the hydrodynamics couples the dry modes within a parity block. The phase now varies continuously with $\kappa$, and near each resonance the dominant mode's phase approaches $\pm 90^{\circ}$, the familiar quarter-cycle lag of a lightly damped oscillator driven at resonance. As $d\to 0$, the coupled prediction converges to its uncoupled counterpart.

\begin{figure}
  \centering
  \includegraphics[width=\textwidth]{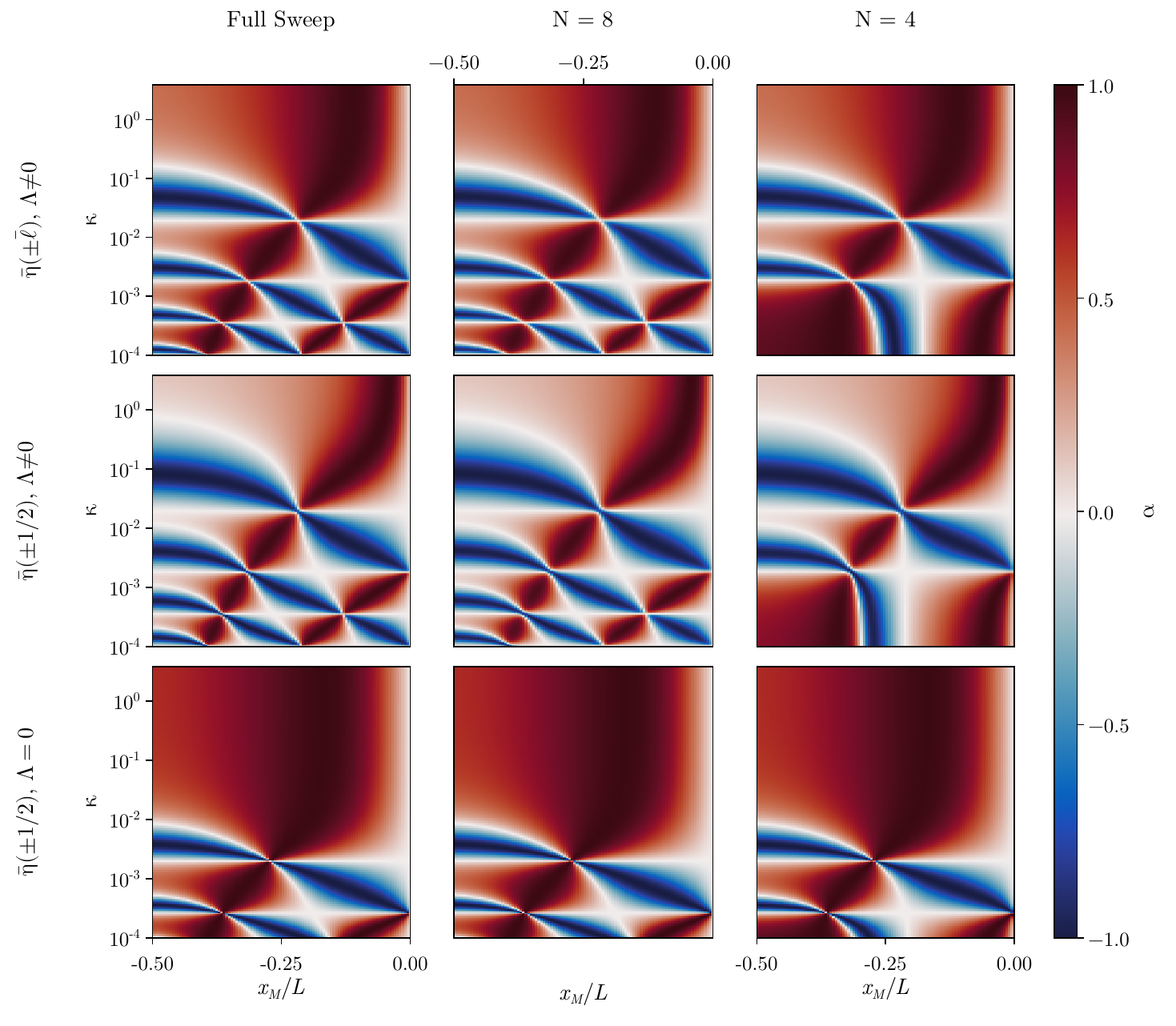}
  \caption{Comparison of full model predictions and modal model reconstructions of the asymmetry factor. Rows compare \(\alpha\) from the far field amplitudes \(\bar{\eta}(\pm\bar{\ell})\) with \(\Lambda\ne0\), the beam end amplitudes \(\bar{\eta}(\pm1/2)\) with \(\Lambda\ne0\), and the beam end amplitudes \(\bar{\eta}(\pm1/2)\) with \(\Lambda=0\). Columns show the full finite difference sweep, the \(N=8\) modal reconstruction, and the \(N=4\) modal reconstruction.}
  \label{fig:modal_maps_3x3}
\end{figure}

The results for the beam endpoint asymmetry for the uncoupled case are presented in the bottom row of figure~\ref{fig:modal_maps_3x3}.  While more quantitative differences are notable, the alternating-block pattern persists.  As such this particular structure is an integral feature of the underlying modal dynamics of the beam, with hydrodynamic feedback shifting and reorganizing the various branches, while also coupling the dry modes. We also note that for the hydrodynamically uncoupled case, the matrix $\mathbf{\bar M}$ becomes diagonal, and we estimate
\begin{align*}
    \bar{q}_n &= \frac{1}{\bar{M}_{nn}} \bar{f}_n = \frac{1}{\bar{M}_{nn}} \int_{-1/2}^{1/2} \bar{f}(x) W_n(x)\,\dd x \\
    &\approx
    \frac{1}{\bar{M}_{nn}} \int_{-1/2}^{1/2} \delta(x-x_M/L)\,W_n(x)\,\dd x = \frac{W_n(x_M/L)}{\bar{M}_{nn}},
\end{align*}
provided the forcing width is small enough, consistent with the idea that a mode cannot be excited when driven at a node. Thus $\bar{q}_n \approx 0$ at the zeros of $W_n$, and the white contour lines intersect exactly at these zeros in figure \ref{fig:modal_maps_3x3}.

The top two rows of figure~\ref{fig:modal_maps_3x3} compare the far-field asymmetry, based on \(\bar{\eta}(\pm\bar{\ell})\), with the beam endpoint asymmetry, based on \(\bar{\eta}(\pm1/2)\) for the fully coupled predictions presented in the main text.  While not identical, the top two rows are qualitatively similar: the alternating-block pattern is similarly present in the beam response itself, before the response is converted into radiated waves. However there are specific regions where the radiated wave asymmetry (top row) and the beam endpoint asymmetry (middle row) are of opposite signs.  This implies that knowledge of the beam endpoint displacements alone is insufficient to predict the direction of net wave thrust.  

Comparing columns in figure~\ref{fig:modal_maps_3x3} also allows us to test the modal truncation. The \(N=8\) reconstruction is visually indistinguishable from the full finite-difference sweep in all three rows. The \(N=4\) reconstruction reproduces the overall alternating-band organization but visibly misses or distorts finer branch structure at small \(\kappa\), most clearly in the top two rows, where the higher modes contribute meaningfully to the full dynamics. Eight modes are therefore the truncation used throughout the rest of this paper; four are shown here only to illustrate that coarser truncations degrade the finer resonance structure at low values of $\kappa$ while preserving the overall qualitative picture.

\subsection{Motor position for asymmetric kinematics in uncoupled rigid limit}\label{app:rigid-optimum}
As a motivating example of the engineering implications that follow from a theoretical analysis in the uncoupled, rigid limit, we present an example where the motor position corresponding to maximal asymmetry can be derived analytically. Let \(X=x_M/L\) denote the non-dimensional forcing position on \([-1/2,1/2]\). In the uncoupled limit and with \(\kappa\to\infty\), the elastic modes are suppressed and the two rigid free--free modes are
\begin{equation}
W_0(x)=1,\qquad W_1(x)=\sqrt{12}\,x .
\end{equation}
For this estimate we model the localized vertical load as \(\bar{f}(x)=F\delta(x-X)\). The forcing projections are therefore
\begin{equation}
\bar{f}_m=\int_{-1/2}^{1/2}F\delta(x-X)W_m(x)\,\dd x=F W_m(X),
\end{equation}
so the two rigid-mode projections are
\begin{equation}
\bar{f}_0=F,\qquad \bar{f}_1=F\sqrt{12}\,X .
\end{equation}
In \eqref{eq:modal_radiation_map}, setting \(\Lambda=0\) decouples \(\bar{\mathbf{Z}}\) and \(\mathbf{\bar C}^\sigma\) from the radiation map. For the rigid modes, \(\beta_0=\beta_1=0\), so their modal equations  \eqref{eq:modal_radiation_map} reduce to \(-\bar{q}_m=-\bar{f}_m\). Thus
\begin{equation}
\frac{\bar{q}_1}{\bar{q}_0}=\frac{\bar{f}_1}{\bar{f}_0}=\sqrt{12}\,X  \implies \bar{q}_1=\sqrt{12} X \bar{q}_0.
\end{equation}
The corresponding endpoint displacements are
\begin{equation}
\bar{\eta}\left(\pm \frac{1}{2}\right)=\bar{q}_0\pm\frac{\sqrt{12}}{2}\bar{q}_1=F \left( 1 \pm 6 X\right).
\end{equation}
For non-zero forcing amplitude, setting one of the endpoint amplitudes to zero yields a prediction for the motor position with the maximum beam endpoint asymmetry.  In particular,
\begin{equation}
\bar{\eta}\left(\pm \frac{1}{2}\right)=0
\quad\iff \quad
1 \pm 6 X = 0
\quad\iff\quad
X= \mp \frac{1}{6}.
\end{equation}
As such, a motor position of $|x_M|/L=1/6$ is anticipated to lead to a maximally asymmetric beam endpoint response in the uncoupled and rigid limit. This numerical value is approached in the large \(\kappa\) limit visible in the bottom row of figure~\ref{fig:modal_maps_3x3}.  This position occurs in the interior of the rigid raft, not simply at the end, confirming that a highly asymmetric kinematic response requires a delicate blend of amplitude and phase of the participating modes, even in the simplest possible scenario illustrated here.  Curiously, the predicted value is close to the experimental value used for the SurferBot ($|x_M|/L=0.12$).  As such, the simple estimate $|x_M|/L=1/6$ may serve as a convenient starting point for iterative design work of near-rigid wave-propelling craft when large wave asymmetry is a central objective.  

\bibliographystyle{jfm}
\bibliography{jfm}

\end{document}